\documentclass[aps,prl,reprint,superscriptaddress]{revtex4-2}
\usepackage{graphicx,color}
\usepackage{amsmath}
\usepackage{bm}
\usepackage{dcolumn}
\usepackage{ifthen}
\usepackage{multirow}
\usepackage{threeparttable}
\usepackage{siunitx}
\usepackage{hyperref}
\hypersetup{backref=true, pdfnewwindow=true, linkcolor=blue, colorlinks=true, anchorcolor=blue, citecolor=blue, filecolor=blue,  menucolor=blue, urlcolor=blue}
\usepackage{lineno}
\begin{document}
\title{Probing the spin polarization of an Anderson impurity}

\author{Mahasweta Bagchi}
\affiliation{II. Physikalisches Institut, Universit\"{a}t zu K\"{o}ln, Z\"{u}lpicher Str. 77, 50937 Cologne, Germany \looseness=-1}
\author{Tfyeche Y. Tounsi}
\affiliation{II. Physikalisches Institut, Universit\"{a}t zu K\"{o}ln, Z\"{u}lpicher Str. 77, 50937 Cologne, Germany \looseness=-1}
\author{Affan Safeer}
\affiliation{II. Physikalisches Institut, Universit\"{a}t zu K\"{o}ln, Z\"{u}lpicher Str. 77, 50937 Cologne, Germany \looseness=-1}
\author{Camiel van Efferen}
\affiliation{II. Physikalisches Institut, Universit\"{a}t zu K\"{o}ln, Z\"{u}lpicher Str. 77, 50937 Cologne, Germany \looseness=-1}
\author{Achim Rosch}
\affiliation{Institut für Theoretische Physik, Universit\"{a}t zu K\"{o}ln, Z\"{u}lpicher Str. 77, 50937 Cologne, Germany \looseness=-1}
\author{Thomas Michely}
\affiliation{II. Physikalisches Institut, Universit\"{a}t zu K\"{o}ln, Z\"{u}lpicher Str. 77, 50937 Cologne, Germany \looseness=-1}
\author{Wouter Jolie}
\affiliation{II. Physikalisches Institut, Universit\"{a}t zu K\"{o}ln, Z\"{u}lpicher Str. 77, 50937 Cologne, Germany \looseness=-1}
\author{Theo A. Costi}
\affiliation{Peter Gr\"{u}nberg Institut, Forschungszentrum Jülich, 52425 J\"{u}lich, Germany \looseness=-1}
\author{Jeison Fischer}
\email{jfischer@ph2.uni-koeln.de}
\affiliation{II. Physikalisches Institut, Universit\"{a}t zu K\"{o}ln, Z\"{u}lpicher Str. 77, 50937 Cologne, Germany \looseness=-1}

\date{\today}

\begin{abstract}
We report spin-polarized scanning tunneling microscopy measurements of an Anderson impurity system in  MoS$_{2}$ mirror twin boundaries, where both the quantum confined impurity state and the Kondo resonance resulting from the interaction with the substrate are accessible. Using a spin-polarized tip, we observe magnetic field induced changes in the peak heights of the Anderson impurity states as well as in the magnetic field-split Kondo resonance. Quantitative comparison with numerical renormalization group calculations provides evidence of the notable spin polarization of the spin-resolved impurity spectral function under the influence of a magnetic field. Moreover, we extract the field and temperature dependence of the impurity magnetization from the differential conductance measurements and demonstrate that this exhibits the universality and asymptotic freedom of the $S=1/2$ Kondo effect. This work shows that mirror twin boundaries can be used as a testing ground for theoretical predictions on quantum impurity models.

\end{abstract}

\maketitle
\newpage

The Kondo effect represents a paradigmatic example where many-body correlations play a pivotal role for its manifestation \cite{Hewson1997}. The microscopic origin of this phenomenon is rooted in  the Anderson impurity model \cite{Anderson1961}, which was originally introduced to describe how magnetic moments form in correlated impurity states that hybridize with a sea of conduction electrons. In the Kondo effect these magnetic moments are gradually screened by the surrounding conduction electrons upon decreasing temperature. While the effect is conceptually straightforward, the theoretical framework to quantitatively describe it is demanding, requiring techniques such as the numerical renormalization group (NRG) \cite{KGWilson1975,KWW1980a} or the Bethe Ansatz \cite{Andrei1983,Kawakami1981,Wiegmann1983a}.

\begin{figure}[h]
\centering
\includegraphics[width=.45\textwidth]{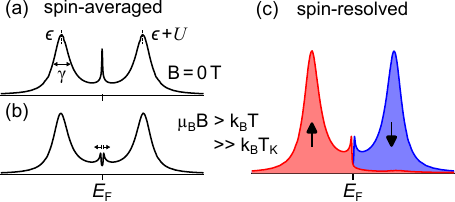}
\caption{
Influence of a magnetic field on the spectral function of a magnetic impurity.
Sketches based on NRG simulation ($\epsilon = -20$\,meV, $U = 40$\,meV, $\gamma = 9.0$\,meV). (a)~At $0$\,T, both spin-components are equally distributed below and above the $E_\mathrm{F}$, resulting in an unpolarized state.
An external field ($g\mu_\mathrm{B}B$, with $g=2.5$) 
splits the Kondo resonance (marked by arrows) 
in the spin-averaged case (b) and induces a massive reorganization of the spectral distribution of the two spin-components in the spin-resolved case (c).
}
\label{fig_intro}
\end{figure}

On the experimental front, scanning tunneling microscopy (STM) and spectroscopy (STS) techniques have stimulated new research on the Kondo effect over the last 25 years \cite{Li1998,Madhavan1998,Wahl2005,Zhao2005,Otte2008,Temirov2008,Zhang2013,Trishin2023,Meng2024}. STM studies typically involve magnetic adsorbates on metallic surfaces \cite{Ternes2017}. Identification of the Kondo effect in spin-averaged STS often relies on  measuring a Kondo zero-bias resonance (ZBR) \cite{Ternes2009}. However, an unambiguous identification of the Kondo effect would ideally require measuring both the Kondo ZBR and the impurity peaks that give rise to the Kondo effect. For adsorbates on metallic surfaces, access to impurity levels has been elusive, due to the dominant presence of substrate states and their strong hybridization to the impurity states. This drawback impedes the exploration of the close connection between the Anderson and Kondo Hamiltonians as established by the Schrieffer-Wolff transformation \cite{Schrieffer1966}. It prevents, for example, identification of the large spectral weight rearrangements predicted for the spin-resolved spectral functions of an Anderson impurity in the presence of a magnetic field, even when the latter exceeds the thermal energy \cite{Hofstetter2000,Costi2001} (see Fig.~\ref{fig_intro}). Previous spin-polarized STM measurements on adsorbates have demonstrated the spin polarization of the split Kondo ZBR \cite{Fu2012,vBergmann2015,Choi2016,Frauhammer2021}, but corresponding measurements of the Anderson model impurity peaks, where most of the spectral weight rearrangement occurs, have been absent.

In this Letter, we present spin-resolved STM measurements on MoS$_2$ mirror-twin boundaries (MTBs), where an Anderson impurity model is realized in a discrete half-filled quantum-confined state, for which both the impurity peaks and the Kondo resonance could be measured \cite{vanEfferen2024}.
With spin-polarized STS, we detect changes in peak intensity of the quantum confined impurity states with varying magnetic field, mirroring the changes seen in the split Kondo resonance. It demonstrates the field dependent spectral weight rearrangement in the spin-resolved spectral functions in quantitative agreement with NRG simulations. By extracting field and temperature dependence of the impurity magnetization from the differential conductance, we demonstrate the universality and asymptotic freedom expected for a $S=1/2$ Kondo system. Asymptotic freedom, a property of the strong interaction between a localized spin and the conduction electrons \cite{Gross1973,Politzer1974}, applies to the $S=1/2$ Kondo model, since here, as in the former, the interaction decreases with increasing energy, field, or temperature scales and significantly impacts properties such as the magnetization.

%Figure 1

We begin by examining the spin-resolved spectral function of a Kondo system as described by the Anderson impurity model \cite{Anderson1961,Hewson1997} (Sec.~V in Supplemental Material \cite{sm}). An impurity state next to the Fermi level splits into a singly occupied state at $\epsilon$ and a doubly occupied state at $\epsilon+U$, owing to the Coulomb repulsion $U$. The impurity state is exchange-coupled to an electron bath, giving rise to the Kondo resonance at the Fermi energy. The hybridization between the impurity and the bath is captured by the width ($\gamma$) of the impurity peak. With the three parameters ($U$, $\epsilon$, $\gamma$) at hand, both impurity states and the Kondo resonance can be accessed through simulation within NRG theory (Sec.~VI in SM \cite{sm}). In the absence of an external magnetic field ($B = 0$\,T), the spin in the occupied impurity state can freely flip [Fig.~\ref{fig_intro}(a)]. Applying an external field $B>0$ affects the spin-averaged spectrum only in the region of the Kondo resonance, causing the latter to split [see Fig.~\ref{fig_intro}(b)]. For the spin-resolved spectral functions, an external field has far-reaching consequences [see Fig.~\ref{fig_intro}(c)]. It shifts almost all the intensity of the up-spin spectral function from the unoccupied peak, far above the Fermi level, to the occupied peak, far below the Fermi level, and the opposite for the down-spin spectral function. Hence, spin-polarized STM should be capable of detecting clear changes in signal intensity in a magnetic field.

The measurements were performed in an ultrahigh vacuum STM operating at $T = 0.35$\,K and equipped with a vector magnetic field up to 9\,T. Spin-polarized STM tips were prepared by coating a W tip with Fe and checked by imaging the spin spiral on Fe islands on Cu(111) \cite{Phark2014}. (for details see Sect. I in SM \cite{sm}). Fe-coated tips display soft magnetization, i.e., their magnetization follows the direction of the external field. Their spin polarization (p) is in the range from 0.2 to 0.4 \cite{Sinkovic1995,Bode1998}. The growth of monolayer MoS$_2$ containing MTBs is achieved on graphene on Ir(111) by Mo deposition in an elemental S pressure of $7 \times 10^{-9}$ mbar, followed by an annealing to 1050\,K in the same S background pressure \cite{Hall2018}.

%Figure 2
\begin{figure}[h]
\centering
\includegraphics[width=.45\textwidth]{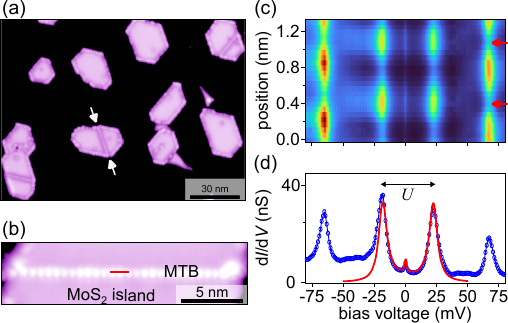}
\caption{Anderson impurity model in MoS$_2$ MTBs.
(a)~STM overview image of single-layer MoS$_2$ islands containing MTBs.
(b)~Atomically resolved image displaying a single MTB with length $L = 18.0$\,nm, indicated by white arrows in (a). %, $\epsilon = -19.0$\,meV, $U = 41.0$\,meV, $\gamma = 7.7$\,meV taken within the rectangle displayed in (a). 
(c)~Differential conductance color plot extracted from linescan along the red line in (b). 
(d)~Differential conductance spectrum (blue circles), obtained after averaged between positions marked by red arrows in (c) and the corresponding NRG simulation (red line; $\epsilon = -18.5$\,meV, $U = 40.5$\,meV, $\gamma = 7.7$\,meV) 
Image information:
 (a)~size $\mathrm{146\,nm\times108}$\,nm, $V_\mathrm{stab} = 1$\,V, $I_\mathrm{stab} = 11$\,pA, $T_\mathrm{s} = 0.4$\,K; 
 (b)~size $\mathrm{20\,nm\times5}$\,nm, $V_\mathrm{stab} = 100$\,mV, $I_\mathrm{stab} = 10$\,pA, $T_\mathrm{s} = 0.4$\,K; 
 (c)~$V_\mathrm{stab} = 100$\,mV, $I_\mathrm{stab} = 1$\,nA, $V_\mathrm{mod} = 0.5$\,mV, $f_\mathrm{mod} = 719$\,Hz, $T_\mathrm{s} = 0.4$\,K.
}
\label{fig_system}
\end{figure}

The Anderson impurity state is realized as a confined MTB state in an MoS$_2$ island, which couples to the underlying electron bath of graphene on Ir(111). An MTB is highlighted by white arrows in Fig.~\ref{fig_system}(a). It exhibits a periodic beating pattern of its confined states close to the Fermi level [Fig.~\ref{fig_system}(b)]. To establish our Kondo system, it is necessary to have an odd number of electrons in the MTB, achieved when a quantum confined state resides and splits around the Fermi energy, a condition induced through controlled charging of the MTB via bias pulses \cite{Yang2022,vanEfferen2024}.

%Figure 3
\begin{figure*}[t]
\centering
\includegraphics[width=\textwidth]{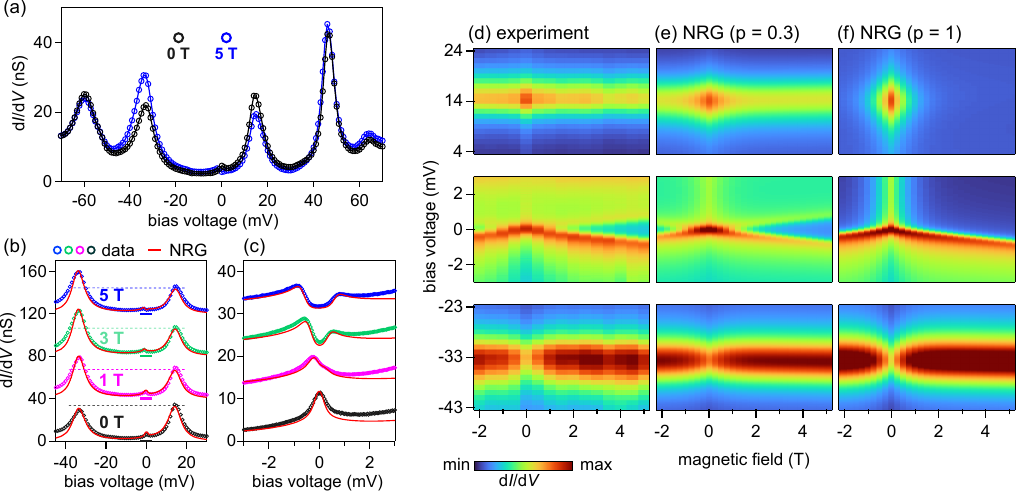}
\caption{Spin-polarized STS data of the Kondo effect in MTBs.
(a)~$\mathrm{d}I/\mathrm{d}V$ spectra with (blue) and without (black) external magnetic field of the impurity peaks and higher order confined states taken with a spin-polarized STM tip.
$\mathrm{d}I/\mathrm{d}V$ spectra of the impurity peaks (b) and Kondo resonance (c) at different magnetic fields taken with the same spin-polarized STM tip (colored circles, MTB with $L = 16.6$\,nm, $\epsilon = -33.0$\,meV, $U = 47.5$\,meV, $\gamma = 7.4$\,meV) and NRG simulation (red lines) using (\ref{didvsp}) with a tip polarization of $\mathrm{p}=0.3$. Dashed lines in (b) are a guide to the change in spectral weight.
(d),(e),(f)~Differential conductance color plots of the unoccupied impurity peak (top panel), Kondo resonance (middle panel), and occupied impurity peak (bottom panel) as a function of bias and magnetic field, comparing the experimental $\mathrm{d}I/\mathrm{d}V$ data (d) and NRG simulations with a tip polarization of $\mathrm{p}=0.3$ (e) and $\mathrm{p}=1$ (f). 
Stabilization parameters: 
 (a),(b)~$V_\mathrm{stab} = 70$\,mV, $I_\mathrm{stab} = 1$\,nA, $V_\mathrm{mod} = 1$\,mV, $f_\mathrm{mod} = 719$\,Hz, $T_\mathrm{s} = 0.4$\,K; 
 (c)~$V_\mathrm{stab} = 5$\,mV, $I_\mathrm{stab} = 1$\,nA, $V_\mathrm{mod} = 100$\,$\mu$V, $f_\mathrm{mod} = 719$\,Hz, $T_\mathrm{s} = 0.4$\,K.
}
\label{fig_sp}
\end{figure*}

To identify the Anderson impurity state, we analyze a series of spin-averaged differential conductance ($\mathrm{d}I/\mathrm{d}V$) spectra taken at $0$\,T on the MTB, as exemplified in Fig.~\ref{fig_system}(c), taken along the red line in Fig.~\ref{fig_system}(b). Focusing on the inner three peaks in Fig.~\ref{fig_system}(d), we identify our Anderson system, comprising the ZBR and the single occupied state below and the doubly occupied state above the Fermi level, all being in phase. These peaks are readily comparable to our spin-averaged NRG simulation (red curve), which uses experimental values for $U$, $\epsilon$, $\gamma$ as input. 
       
We now explore how the impurity peaks and the ZBR respond to an external magnetic field using spin-polarized STM. Fig.~\ref{fig_sp}(a) shows differential conductance spectra ($\mathrm{d}I/\mathrm{d}V$) at $0$\,T and under external magnetic field of $5$\,T normal to the sample surface (spin-polarized measurements of MTB in Fig.~\ref{fig_system} shown in Fig.~S3 in SM \cite{sm}). The external field increases the intensity of the occupied impurity peak and the unoccupied peak diminishes, rendering them asymmetric, while the outer, higher energy peaks remain unchanged. 
From the sequences of $\mathrm{d}I/\mathrm{d}V$ spectra at various fields for the impurity peaks [Fig.~\ref{fig_sp}(b)], the continuous change in asymmetry is observed. The field also affects the ZBR [Fig.~\ref{fig_sp}(c)], it is Zeeman split \cite{Rosch2003} and a spin-excitation step-like increase centered at $E_\mathrm{F}$ appears. The split ZBRs are also asymmetric in clear correlation with the impurity peaks, i.e., occupied states enhanced and unoccupied states diminished. The field dependence in Fig.~\ref{fig_sp}(b) and~\ref{fig_sp}(c) are due to the spin filter effect caused by the spin-polarized STM tip, since the $\mathrm{d}I/\mathrm{d}V$ signal acquires a dependence on the spin polarization of the tip and the spin-polarized spectral functions of the sample, see Eq.~(S9) in \cite{sm}.
The field induced changes seen in Fig.~\ref{fig_sp}(a) is not observed with a spin-averaged tip ($\mathrm{p}=0$), as the intensity of impurity peaks is unchanged, see Fig.~S2 in \cite{sm}. The in-field asymmetry of the impurity states is evidence of the spin-polarized nature of the states and is a direct result of the spectral weight redistribution due to the external field, visualized in Fig.~\ref{fig_intro}(c).

To quantify the changes in terms of the spin polarization, we compare our experimental data to field dependent $\mathrm{d}I/\mathrm{d}V$ spectra simulated within the framework of NRG.
At low temperatures, the spin polarized differential conductance $\mathrm{d}I/\mathrm{d}V$ for the Anderson model is given by (see \cite{sm}),
\begin{equation}
\frac{\mathrm{d}I}{\mathrm{d}V} \propto [(1+\mathrm{p})/2]\frac{\mathrm{d}I}{\mathrm{d}V}_\uparrow+[(1-\mathrm{p})/2]\frac{\mathrm{d}I}{\mathrm{d}V}_\downarrow,
\label{didvsp}
\end{equation}
where $\mathrm{d}I/\mathrm{d}V_i=(4/\pi \gamma)A_i$, and $A_{i=\uparrow,\downarrow}$, are the energy, temperature and magnetic field dependent spectral functions of the Anderson model and $\mathrm{p}$ is the polarization of the tip. The simulated $\mathrm{d}I/\mathrm{d}V$ is adjusted to match the experimental data at 5\,T resulting in an estimated $\mathrm{p}=0.3$, with the simulation displayed as red lines in the Fig.~\ref{fig_sp}(b) and~\ref{fig_sp}(c).

The NRG simulated $\mathrm{d}I/\mathrm{d}V$, obtained with Eq.~\ref{didvsp} with the extracted $\mathrm{p}$, for different external fields is displayed in Fig.~\ref{fig_sp}(b) and \ref{fig_sp}(c). The simulation resembles the experimental data quite accurately. The agreement is substantiated by comparing the experimental and NRG data color plots for an extended range of magnetic field displayed in Fig.~\ref{fig_sp}(d) and ~\ref{fig_sp}(e). In both experiment and theory, the asymmetry between unoccupied (top panels) and occupied (bottom panels) impurity peaks increases as a function of field and reaches saturation at about 2\,T, while the ZBR in the middle panels transforms into a gap with pronounced larger occupied peak at larger fields. The agreement obtained simultaneously for impurity peaks and Kondo resonance with NRG by just introducing a tip spin polarization is unambiguous evidence that the impurity peaks are part of the Kondo system.
The effect of a magnetic field on the spin polarization of the MTB can best be visualized by artificially setting $\mathrm{p}=1$, i.e., by assuming that the tip acts as a perfect spin filter.
In this case, as shown in Fig.~\ref{fig_sp}(f), only the occupied peak is present at the highest field, while the unoccupied peak is completely suppressed [see also Fig.~\ref{fig_intro}(c)].

\begin{figure}[h]
\includegraphics[width=0.45\textwidth]{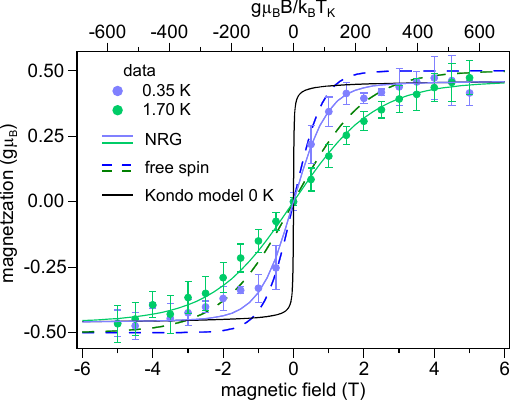}
\caption{Field dependence of the magnetization as extracted from the measured asymmetry using (\ref{eq:mag-asymmetry-link}) for the data in Fig.~\ref{fig_sp} (blue circles) with $\mathrm{p}=0.3$, and for the data obtained at $1.70$\,K presented in the Fig.~S4 \cite{sm} for the same MTB (green circles). 
NRG thermodynamic magnetization for the parameters: $\epsilon = -33.0$\,meV, $U = 47.5$\,meV, $\gamma = 7.4$\,meV. Dashed lines: magnetization for a free spin $1/2$. Black solid line: $T=0$ Kondo model magnetization. For a comparable plot featuring a different MTB, see Fig.~S5 in \cite{sm}.
}
\label{fig_temp}
\end{figure}

The fact that we can align the magnetic moment using a magnetic field - in analogy to an isolated paramagnetic spin - seems to contradict the Kondo effect, which screens the magnetic moment and thus forms a non-magnetic spin-singlet (at $T=0$). However, even at $T=0$, this many-body singlet is polarizable. At finite temperature,
information on the polarizability of the Kondo system is contained in the thermodynamic magnetization $m(B,T)$ of the Anderson impurity model describing the MTB (see Sec.~VIII in \cite{sm}). We can relate the latter to the measured field and temperature dependence of the $\mathrm{d}I/\mathrm{d}V$-weight-asymmetry, denoted by $\mathrm{A}_{\mathrm{d}I/\mathrm{d}V}^w$, and defined by
\begin{equation}
\mathrm{A}_{\mathrm{d}I/\mathrm{d}V}^w(B,T)=\frac{\mathrm{d}I/\mathrm{d}V_{a}(B,T)-\mathrm{d}I/\mathrm{d}V_{b}(B,T)}{\mathrm{d}I/\mathrm{d}V_{a}(B,T)+\mathrm{d}I/\mathrm{d}V_{b}(B,T)},
\label{eq_a}
\end{equation}
where $\mathrm{d}I/\mathrm{d}V_{a}$ and $\mathrm{d}I/\mathrm{d}V_{b}$ are the $\mathrm{d}I/\mathrm{d}V$ weights above and below the Fermi level, respectively. Specifically, $m(B,T)$ is related to $\mathrm{A}_{\mathrm{d}I/\mathrm{d}V}^w(B,T)$ and the spin polarization of the tip ($\mathrm{p}$) via (see \cite{sm}), 
\begin{equation}
  m(B,T)=  -g\mu_B\left(\mathrm{A}_{\mathrm{d}I/\mathrm{d}V}^w(B,T)-\mathrm{A}_{\mathrm{d}I/\mathrm{d}V}^w(0,T)\right)
  /2\mathrm{p}.\label{eq:mag-asymmetry-link}
\end{equation}
Results for the magnetization of MTB obtained in this way, at $0.35$\,K (blue circles) and at $1.70$\,K (green circles) are shown in Fig.~\ref{fig_temp}.
Also shown are the corresponding magnetizations for the Anderson model within NRG (blue/green solid lines) and for a free spin $1/2$ (blue/green dashed lines).

Inspection of Fig.~\ref{fig_temp} reveals several interesting aspects. First, the experimental data follows the universal magnetization curve of the Anderson model at each temperature. Universality of the magnetization $m(B,T)$ in the Anderson model means that
for each fixed $T$ (or equivalently each fixed $T/T_K$), $m(B,T)$ is a universal function of $g\mu_{B}B/k_BT_K$ (whose form changes continuously 
upon varying $T/T_K$ from $0$ to $\gg 1$, see Fig.~S12 in \cite{sm}). Second, the measured magnetization, characterized by a linear increase at low fields, only gradually saturates at higher fields. This behavior is in stark contrast to the magnetization curve of a free spin $1/2$. The latter saturates rapidly to the fully polarized value $\pm 1/2$ at high fields (dashed lines). The magnetization of the MTB remains lower even at high fields and approaches the fully polarized value only slowly following the NRG magnetization which, for sufficiently high fields, exhibits the theoretically well known logarithmic corrections. This property of the magnetization is a characteristic of the asymptotic freedom in the Kondo effect \cite{Andrei1983}, resulting in the observed diminished magnetization at high fields. This behaviour, most markedly illustrated by the $T=0$ Kondo model magnetization from the Bethe Ansatz \cite{Andrei1982} (black solid line), is eventually cut off on field scales $g\mu_B B \sim (U\gamma)^{1/2}$ in the case of the Anderson model \cite{Wiegmann1983a} (for details see Secs.~VIII.A-B \cite{sm}). Above this scale, of order $10-20$ meV for our MTBs, the fully polarized (free spin) value would eventually be restored (Figs.~S7 and S8 \cite{sm}), however such large fields ($100-200$~T) are experimentally inaccessible. The field range used in the experiment is therefore accessing the universal part of the magnetization curve of the Anderson model (see Figs.~S7 and S8 in \cite{sm}). Finally, the variation of the magnetization with temperature is also quantitatively well described by the NRG, using just the experimentally measured parameters $U,\epsilon$ and $\gamma$.

To conclude, our spin-polarized STM experiments show that the spin-resolved impurity peaks that give rise to a Kondo resonance undergo a massive spectral weight redistribution under external magnetic fields exceeding the thermal energy involved, meaning that the peak below $E_\mathrm{F}$, i.e., occupied state, is composed of majority spin state and the peak at positive bias voltage, i.e., unoccupied state, is dominated by minority spin state. This description is in full agreement with the Anderson model, tackled in the framework of NRG. Analysis of the magnetization response of the MTB with respect to field and at different temperatures sheds light on the spin $1/2$ character of the system, and reveals the typical signatures of asymptotic freedom characteristic of the Kondo effect, in marked contrast to a free spin $1/2$. MTBs provide a versatile realization of the standard model of correlated electrons, the Anderson impurity model, allowing a wide range of Kondo scales to be realized through control of the model parameters $U,\epsilon$ and $\gamma$ \cite{vanEfferen2024}. This, together with spin-polarized STM opens up new prospects for controllably investigating universal aspects of strongly correlated systems, including investigations of multi-impurity systems \cite{Jones1988,Ingersent2005} using MTBs or magnetic MTBs in the presence of superconductivity.
\\
\begin{acknowledgments}
	\noindent
	We acknowledge funding from Deutsche Forschungsgemeinschaft (DFG) through CRC 1238 (project No. 277146847, subprojects B06 and C02). W.J. acknowledges financial support from the DFG through project JO 1972/2-1 (project No. 535290457) within the SPP 2244. T.A.C. gratefully acknowledges computing time on the supercomputer JURECA \cite{JURECA} at Forschungszentrum Jülich under grant no. JIFF23. J.F. acknowledges financial support from the DFG through project FI 2624/1-1 (project No. 462692705) within the SPP 2137.
\end{acknowledgments}

\bibliographystyle{apsrev4-2}
\bibliography{bib_sp-Kondo}
\end{document}

\title{Supplemental Material:\\%Supporting Information:\\
Probing the spin polarization of an Anderson impurity}

\author{Mahasweta Bagchi}
\affiliation{II. Physikalisches Institut, Universit\"{a}t zu K\"{o}ln, Z\"{u}lpicher Str. 77, 50937 Cologne, Germany \looseness=-1}
\author{Tfyeche Y. Tounsi}
\affiliation{II. Physikalisches Institut, Universit\"{a}t zu K\"{o}ln, Z\"{u}lpicher Str. 77, 50937 Cologne, Germany \looseness=-1}
\author{Affan Safeer}
\affiliation{II. Physikalisches Institut, Universit\"{a}t zu K\"{o}ln, Z\"{u}lpicher Str. 77, 50937 Cologne, Germany \looseness=-1}
\author{Camiel van Efferen}
\affiliation{II. Physikalisches Institut, Universit\"{a}t zu K\"{o}ln, Z\"{u}lpicher Str. 77, 50937 Cologne, Germany \looseness=-1}
\author{Achim Rosch}
\affiliation{Institut für Theoretische Physik, Universit\"{a}t zu K\"{o}ln, Z\"{u}lpicher Str. 77, 50937 Cologne, Germany \looseness=-1}
\author{Thomas Michely}
\affiliation{II. Physikalisches Institut, Universit\"{a}t zu K\"{o}ln, Z\"{u}lpicher Str. 77, 50937 Cologne, Germany \looseness=-1}
\author{Wouter Jolie}
\affiliation{II. Physikalisches Institut, Universit\"{a}t zu K\"{o}ln, Z\"{u}lpicher Str. 77, 50937 Cologne, Germany \looseness=-1}
\author{Theo A. Costi}
\affiliation{Peter Gr\"{u}nberg Institut, Forschungszentrum Jülich, 52425 J\"{u}lich, Germany\looseness=-1}
\author{Jeison Fischer}
\email{jfischer@ph2.uni-koeln.de}
\affiliation{II. Physikalisches Institut, Universit\"{a}t zu K\"{o}ln, Z\"{u}lpicher Str. 77, 50937 Cologne, Germany \looseness=-1}

\date{\today}
\maketitle
\vspace{-1cm}
\tableofcontents
\newpage
\section{Spin-polarized tip characterization}

To ensure that a spin-polarized tip is used for the STS measurements of the MoS$_2$ mirror twin boundaries (MTBs), Fe-coated STM tips were characterized in a well known magnetic system. The tips were prepared by depositing a thick layer ($20$~ML) of Fe onto a flashed W tip, followed by a mild annealing \cite{Phark2013}. Each tip was calibrated using the spin spiral of bilayer Fe on Cu(111) \cite{Phark2014}. Figure~S1 shows $\mathrm{d}I/\mathrm{d}V$ maps taken at an out-of-plane magnetic field of 2~T on Fe bilayers on Cu(111) with the same tip as in Fig.\,3 (main text) in (a) and in Fig.~\ref{fig_sp_mtb2} in (b). The $\mathrm{d}I/\mathrm{d}V$ maps display the characteristic modulation pattern due to the presence of a spin spiral. The bright and dark modulation stripes appear due to the projection of the rotating sample magnetization onto the out-of-plane tip magnetization orientation.
The strong contrast qualifies the tips for subsequent use with MoS$_2$ MTBs.
%Figure S1
\begin{figure*}[h]
%\centering
\includegraphics[width=\textwidth]{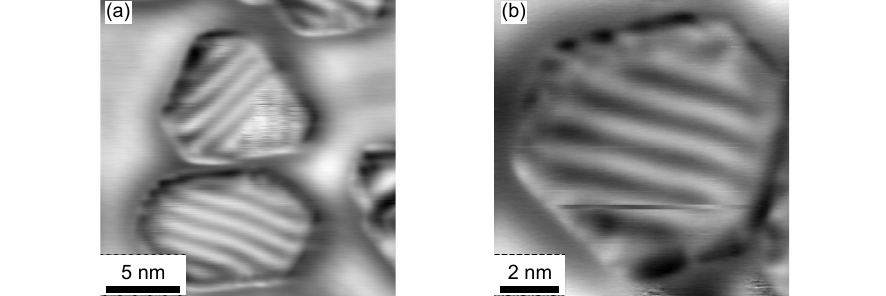}
\caption{Characterization of spin-polarized tips before MTB measurements. Visualization of the spin-spiral in bilayer Fe islands on Cu (111) under external out of plane field of 2~T. (a) $\mathrm{d}I/\mathrm{d}V$ map taken with tip used for measurements displayed in Fig.\,3 (main text) and (b) $\mathrm{d}I/\mathrm{d}V$ map taken with tip used for measurements in Fig.\,S3.
Measurements parameters: (a,b) $V_\mathrm{bias} = -600$\,mV, $I = 1$\,nA, $T_\mathrm{s} = 1.7$\,K. 
}
\label{si_fig_tip}
\end{figure*}

%\newpage
%Figure S2
%\begin{figure}[h]
%\centering
%\includegraphics[width=0.45\textwidth]{figures/NRG_diff_gamma.pdf}
%\caption{This placeholder is to consider including an example of NRG simulation with different gammas, below and above Fermi. }
%\label{si_fig_nonmagtip}
%\end{figure}

\newpage
\section{Measurements with a spin-averaged tip}
%Figure S2
\begin{figure}[h]
\centering
\includegraphics[width=0.5\textwidth]{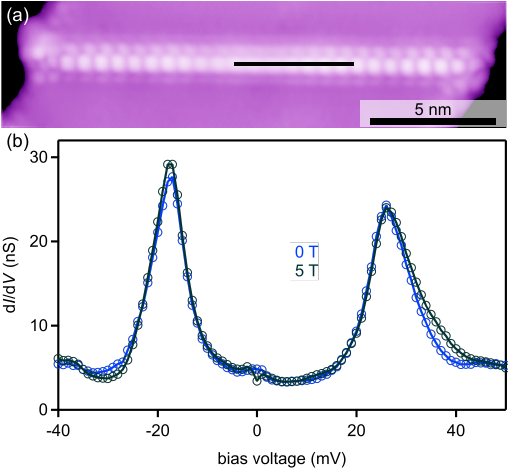}
\caption{Topography displaying an MTB and spin-averaged STS data of the Kondo system measured with a non-magnetic tip. (a) atomically resolved image displaying a single MTB. (b) Spin-average of a differential conductance spectrum measured along the black line on MTB, 0 T in blue and 5T in black. Measurements parameters: (a) $V_\mathrm{stab} = 100$\,mV, $I_\mathrm{stab} = 20$\,pA, $T_\mathrm{s} = 0.4$\,K; 
 (b) $V_\mathrm{stab} = 100$\,mV, $I_\mathrm{stab} = 10$\,pA, $T_\mathrm{s} = 0.4$\,K;(b) ~$V_\mathrm{stab} = 100$\,mV, $I_\mathrm{stab} = 1$\,nA, $V_\mathrm{mod} = 1$\,mV, $f_\mathrm{mod} = 719$\,Hz, $T_\mathrm{s} = 0.4$\,K. 
}
\label{si_fig_nonmagtip}
\end{figure}

\newpage
\section{Spin-polarized data for a different MTB}
%Figure S3
\begin{figure*}[h]
%\centering
\includegraphics[width=\textwidth]{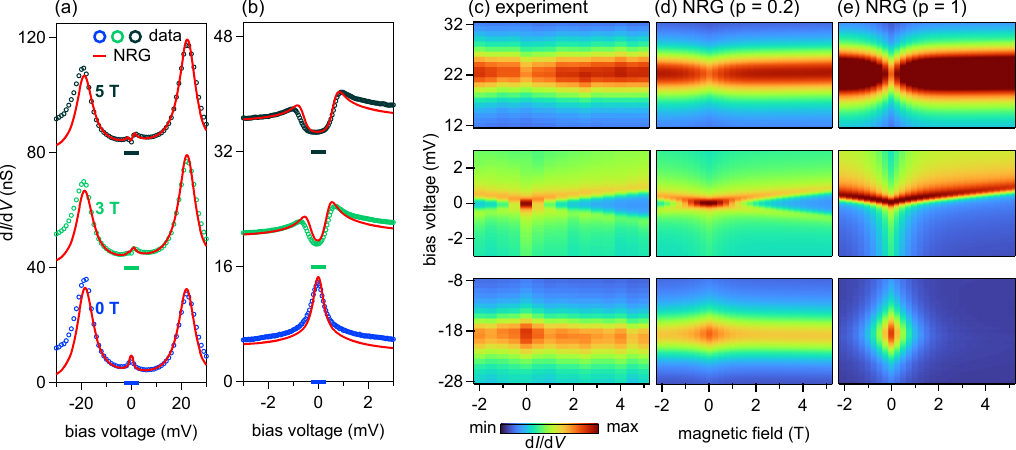}
\caption{Spin-polarized STS data of the Kondo effect in MTB2 (same as Fig.\,2 in the main).
$\mathrm{d}I/\mathrm{d}V$ spectra of the impurity peaks (a) and Kondo resonance (b) at different magnetic fields taken with a spin-polarized STM tip (colored circles) and NRG calculation (red line) with convoluted up and down components for a tip spin polarization of $\mathrm{p}=0.2$.
(c),(d),(e)~Differential conductance color plots of the unoccupied impurity peak (top range), Kondo resonance (middle range), and occupied impurity peak (bottom range) as a function of bias and magnetic field, comparing the experimental $\mathrm{d}I/\mathrm{d}V$ data (c) and NRG simulations with a tip spin polarization of $\mathrm{p}=-0.2$ (d) and $\mathrm{p}=-1$ (e). %Color scale of upper and lower panel of (c),(d),(e) = 0 to 66 nS; color scale of middle panel of (c),(d),(e) = 3 to 14 nS.
Stabilization parameters: 
 (a)~$V_\mathrm{stab} = 100$\,mV, $I_\mathrm{stab} = 1$\,nA, $V_\mathrm{mod} = 0.5$\,mV, $f_\mathrm{mod} = 719$\,Hz, $T_\mathrm{s} = 0.4$\,K; 
 (b)~$V_\mathrm{stab} = 5$\,mV, $I_\mathrm{stab} = 1$\,nA, $V_\mathrm{mod} = 100$\,$\mu$V, $f_\mathrm{mod} = 719$\,Hz, $T_\mathrm{s} = 0.4$\,K.
 }
\label{fig_sp_mtb2}
\end{figure*}

Note: The data displayed here is referred to MTB2, same MTB displayed in Fig.~2 (main text). The spin-polarized data displayed in Fig.~3 refers to MTB1. Their details are given in Tab.~\ref{Table1}
\newpage
\section{Spin-polarized data at 1.7 K}
%Figure S4
\begin{figure*}[h]
%\centering
\includegraphics[width=\textwidth]{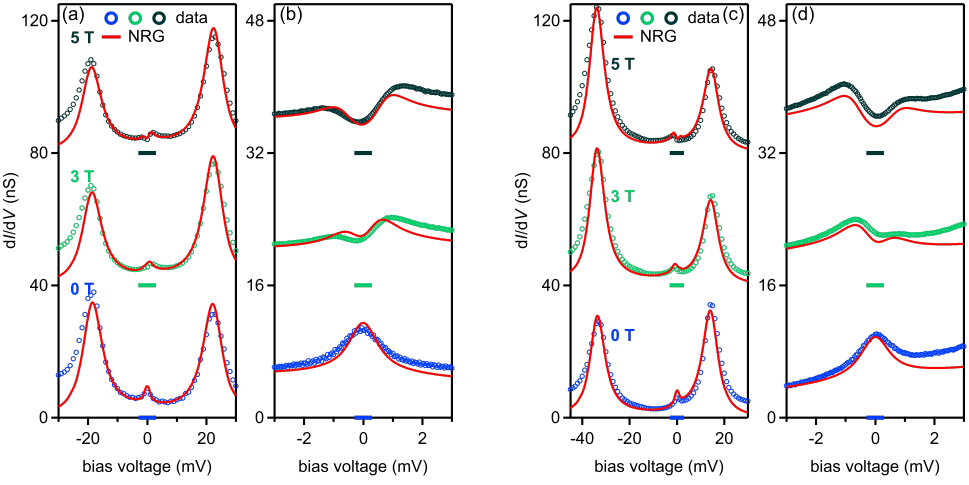}
\caption{Spin-polarized STS data of the Kondo effect in MTBs at higher temperature of $1.7$\,K, instead of $0.35$\,K.
$\mathrm{d}I/\mathrm{d}V$ spectra of the impurity peaks (a) and Kondo resonance (b) at different magnetic fields taken with a spin-polarized STM tip (colored circles) on MTB2, same as Fig.~S3 and NRG calculation (red line) for a tip spin polarization of $\mathrm{p}=-0.2$. $\mathrm{d}I/\mathrm{d}V$ spectra of the impurity peaks (c) and Kondo resonance (d) at different magnetic fields taken with a spin-polarized STM tip (colored circles) on MTB1, same as Fig.~3 (main text). NRG calculation (red line) for a tip spin polarization of $\mathrm{p}=0.3$.
}
\label{si_fig_NRG_1.7K}
\end{figure*}

\newpage
\section{Magnetization of a different MTB}
%Figure S5
\begin{figure*}[h]
%\centering
\includegraphics[width=0.6\textwidth]{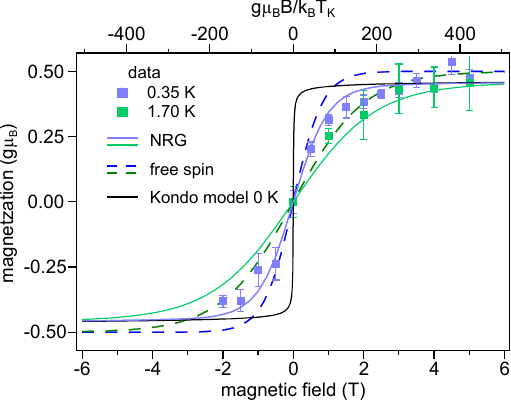}
\caption{Field dependence of the magnetization as extracted from the measured asymmetry, using Eq.~(2) in the main text, for MTB2, data in Fig.~\ref{fig_sp_mtb2} (blue squares) with $\mathrm{p}=-0.2$, and for the data obtained at $1.70$\,K presented in the Fig.~\ref{si_fig_NRG_1.7K} for the same MTB (green squares). 
NRG thermodynamic magnetization for the parameters: $\epsilon = -18.5$\,meV, $U = 40.5$\,meV, $\gamma = 7.7$\,meV. Dashed lines: magnetization for a free spin $1/2$.
}
\label{fig:MTB2-magnetization}
\end{figure*}

\newpage
\section{Model and spin polarized current}
\label{sec:model}   
The NRG calculations of the spin-resolved $\mathrm{d}I/\mathrm{d}V$ spectra of the MTB's are based on the single-level Anderson impurity model in the presence of a local magnetic field $B$ \cite{Anderson1961}.
The experimental set up consists of the impurity, $H_{imp}$, describing the MTB hybridizing with the substrate, a ferromagnetic (superparamagnetic) tip, $H_{tip}$, and a term, $H_{tunnel}$, describing the tunneling of electrons from the edge of the tip to the impurity level. Thus, the system is described by the Hamiltonian,
\begin{align}
  H &= H_{imp}+H_{tip}+H_{tunnel},
\end{align}
with
\begin{align}
  H_{imp}= & \sum_{\sigma}(\epsilon -g\mu_BB\sigma/2) n_{\sigma} +Un_{\uparrow}n_{\downarrow}\nonumber\\  + &\sum_{k\sigma}(\varepsilon_{k}-g'\mu_BB\sigma/2)c_{k\sigma}^{\dagger}c_{k\sigma}
  + 
       \sum_{k\sigma}V_{k}(c_{k\sigma}^{\dagger}d_{\sigma} +d_{\sigma}^{\dagger}c_{k\sigma}),\label{eq:ham}\\
             H_{tip}& = \sum_{k\sigma}(E_{k\sigma}+eV)a_{k\sigma}^{\dagger}a_{k\sigma},\label{eq:tip}\\
             H_{tunnel}&=t_d\sum_{\sigma}(A_{\sigma}^{\dagger}d_{\sigma}+d^{\dagger}_{\sigma}A_{\sigma}).\label{eq:tunnel}
\end{align}
In  (\ref{eq:ham}), $\epsilon$ is the energy of a spin degenerate level, which hybridizes with the substrate conduction electron states (with kinetic energies $\varepsilon_k$) via hybridization matrix elements  $V_{k}$, $n_{\sigma}=d^{\dagger}_{\sigma}d_{\sigma}$ is the occupation number for spin $\sigma={\uparrow,\downarrow}$ electrons in the impurity level, $U$ is a local Coulomb repulsion and $B$ is a magnetic field acting on the impurity electrons via a Zeeman shift $-g\mu_BB\sigma/2$ with $g$ the impurity g-factor. A magnetic field acting on the substrate conduction electrons in Eq.~(\ref{eq:ham}) can be neglected. In the wide band limit, which applies to our system (see below), such a term gives a negligible contribution to impurity properties \cite{Hewson1997,Merker2012b}.
The tip Hamiltonian, $H_{tip}$, with spin-split bands $E_{k\uparrow}\neq E_{k\downarrow}$, has its chemical potential raised by $eV$ relative to the substrate chemical potential which is fixed at $\mu=0$. The term, $H_{tunnel}$, describes tunneling between a local state $A_{\sigma}^{\dagger}|0\rangle$ at the edge of the tip to the localized state of the impurity \cite{Tersoff1985}. The use of an energy independent tunnel matrix element $t_d$ is here justified by the low energy phenomena being investigated.

For weak tunneling $t_d\ll V_{k}$,  the current $I(V)$ to second order in $t_d$ is given by \cite{Schiller2000}
\begin{align}
  I(V) = \sum_\sigma I_{\sigma} = \frac{2e}{\pi\hbar}|t_d|^2\sum_{\sigma}\int_{-\infty}^{+\infty}(f(E-eV)-f(E))N_{\sigma}(E-eV) A_{\sigma}(E)dE, 
\end{align}
where $N_{\sigma}(E)=\sum_{k}\delta(E-E_{k\sigma})$ is the spin-dependent density of states of the ferromagnetic tip and $A_{\sigma}(E)$ is the spin-dependent spectral function of the local impurity state (it depends also on field $B$ and temperature $T$, but for simplicity of notation, this dependence is not explicitly shown).
Thus the current can be calculated from the spin-dependent spectral function of (\ref{eq:ham}) and a knowledge of the spin polarization
of the ferromagnetic tip. We further assume that the tip density of states $N_{\sigma}(E)$ has a negligible energy dependence on the scale of the impurity physics (set by
$U,\epsilon$ and the hybridization strength), but that it has a finite polarization $p=(N_{\uparrow}-N_{\downarrow})/(N_{\uparrow}+N_{\downarrow})=(N_{\uparrow}-N_{\downarrow})/N_F$, where $N_F=N_{\uparrow}+N_{\downarrow}$ is the tip density of states at the Fermi level. The tip used in the experiment is superparamagnetic: it has a finite polarization $p$ for fields $|B|$ exceeding a polarizing field $B_{pol}>0.5$~T, while for $|B|=0$ the tip polarization is zero. Correspondingly, $N_F$ is understood below to take two values: one value $N_F^< $ for $|B|=0$,
and another value $N_F^{>}$ for  $|B|>B_{pol}$. Then, the differential conductance $\mathrm{d}I/\mathrm{d}V$ can be written as
\begin{align}
  \frac{dI}{dV} & = \frac{2e^2}{\pi\hbar}|t_d|^2\int_{-\infty}^{+\infty}dE\left(-\frac{\partial f(E-eV)}{\partial E}\right) \left[N_{\uparrow} A_{\uparrow}(E)+N_{\downarrow} A_{\downarrow}(E)\right], \\
  & =  \frac{2e^2}{\pi\hbar}|t_d|^2N_{F}^{<,>}\int_{-\infty}^{+\infty}dE\left(-\frac{\partial f(E-eV)}{\partial E}\right) \left[\frac{1+p}{2} A_{\uparrow}(E)+\frac{1-p}{2} A_{\downarrow}(E)\right], \label{eq:spin-resolved-dIdV}\\
  & =  c^{<.>}\left[\frac{1+p}{2} A_{\uparrow}(E=eV)+\frac{1-p}{2} A_{\downarrow}(E=eV)\right],\;\;\text{for $T\to 0$.}\label{eq:spin-resolved-dIdV-lowTemp}
\end{align}
where $c^{<,>}=\frac{2e^2}{\pi\hbar}|t_d|^2N_{F}^{<,>}$ are material dependent prefactors for $|B|<B_{pol}$, and $|B|>B_{pol}$.
The expression (\ref{eq:spin-resolved-dIdV})  includes both the spin averaged contribution [proportional to $A_0=\frac{1}{2}(A_\uparrow+A_\downarrow)$] and the spin-polarized contribution [proportional to $A_{sp}=\frac{1}{2}(A_\uparrow-A_\downarrow)$]. It may be written in terms of $A_0, A_{sp}$ and the tip polarization $p$ as,
\begin{align}
  \frac{dI}{dV}   & =  c^{<,>}\int_{-\infty}^{+\infty}dE\left(-\frac{\partial f(E-eV)}{\partial E}\right) \left[A_0(E) + pA_{sp}(E)\right]. 
\end{align}

In the following, we take a constant $V_k=V_0$ for the Anderson impurity model (\ref{eq:ham}) and a constant substrate conduction electron density of states $\rho(E)=\rho =1/2D$ with $-D\leq E \leq +D$, where $D$ is the half-bandwidth. Hence, the hybridization function $\gamma_0(E)=2\pi\sum_{k}|V_k|^2\delta(E-\varepsilon_k)$ is a constant $\gamma_0(E)=\gamma_0=2\pi\rho V_0^2=2\Delta_0$ and equals the  full-width at half-maximum (FWHM) of the $U=0$ resonant level. A note on notation:  we have introduced $\Delta_0=\gamma_0/2$ for the HWHM of the noninteracting resonant level. This is usual in  theoretical studies of the Anderson impurity model \cite{Hewson1997}. In experimental work on quantum dots and adatoms, the full width $\gamma_0$ is generally used. The model is fully characterized by the parameters $U$, $\epsilon$ and $\gamma_0$. 
For the MTB's of interest, the Graphene monolayer on the Iridium substrate is doped and has a metallic density of states at $E_F$ with the Dirac point lying far above \cite{Jolie2019}. The Hubbard excitations at $\epsilon$ and $\epsilon+U$, also termed filled and empty impurity states in the main text,  also lie far from the Dirac point. Thus, to a first approximation, NRG calculations in the wide-band limit of the Anderson impurity model, i.e., $D\gg \text{max}(U, |\epsilon|,|\epsilon+U|,\gamma_0)$, are justified.

For weak correlations, $U/\pi\Delta_0 \lesssim 1$, the physics of the Anderson model is well approximated by a (Lorentzian) resonant level at $E=\epsilon$ with an unrenormalized width $\gamma_0=2\Delta_0$, i.e., the impurity spectral density is given by $A(E,T)=(\Delta_0/\pi)/((E-\epsilon)^2+\Delta_0^2)$. The MTB's considered in this Letter are in the opposite regime of strong correlations, i.e., $U/\pi\Delta_0 \gg 1$. The impurity state is also approximately half-filled, $n_0=\sum_{\sigma}\langle n_{\sigma}\rangle\approx 1$, with $-\epsilon\gg \Delta_0=\gamma_0/2$ and $(\epsilon+U)\gg \Delta_0=\gamma_0/2$ such that 
a $S=1/2$ local moment nominally forms on the impurity (MTB). This impurity spin interacts antiferromagnetically via a Kondo exchange with the surrounding conduction electrons of the substrate resulting in a gradual screening of the impurity spin with decreasing temperature. Eventually,  in the absence of an external magnetic field, a many-body singlet groundstate forms at $T=0$ whose signature in the impurity spectral function is the Kondo resonance pinned close to the Fermi level and having a characteristic width $k_BT_K$ determined by the bare model parameters $U, \epsilon$ and $\Delta_0=\gamma_0/2$ and
given explicitly by
\begin{align}
  k_BT_K & = \gamma_0 \sqrt{U/4\gamma_0}e^{-\pi|\varepsilon||\varepsilon+U|/U\gamma_0}.\;\;\;\text{for $\epsilon/\gamma_0 \ll -1, (\epsilon+U)/\gamma_0\gg +1$ and  $U/\gamma_0\gg 1$}.\label{eq:t0-asymm}
\end{align}
The Kondo effect also significantly modifies the high energy impurity excitations at $\epsilon$ and $\epsilon+U$. Namely, it renormalizes their width $\gamma$
by a factor of two as compared to the bare width $\gamma_0$, i.e., $\gamma=2\gamma_0=4\Delta_0$. Logan et al. \cite{Logan1998} give a simple physical explanation of this: in the Kondo regime an electron in the impurity level can tunnel to the conduction band with or without a spin-flip, resulting in two decay channels and thereby a doubled width for the impurity states. This renormalization is also confirmed by NRG calculations \cite{vanEfferen2024}, which also show that the Hubbard excitations remain Lorentzian in lineshape. The renormalization by a factor of two holds in the symmetric Kondo regime, and deviates slightly from two in the asymmetric Kondo regime. 

The ability to measure directly the impurity states in the MTB's in this Letter allows all the bare model parameters $\epsilon, U$ and $\gamma_0=\gamma/2$ of the Anderson model to be extracted directly from experiment and allows an ab-initio determination of $T_K$ from (\ref{eq:t0-asymm}). These parameters, together with the experimental temperature and field values, then allow for a quantitative modeling of the spin polarized $\mathrm{d}I/\mathrm{d}V$ measurements on all relevant bias voltage scales by using the NRG technique. For reference, we list in Table~\ref{Table1} the extracted model parameters and ab-initio determined Kondo scales for the two MTBs investigated in the present work.
\begin{table}[h]
\begin{ruledtabular}
  \begin{tabular}{ccccc}
   \multicolumn{1}{c}{MTB} & 
   \multicolumn{1}{c}{$U$ ({\rm meV} }) & 
   \multicolumn{1}{c}{$\epsilon$ ({\rm meV}) }& 
   \multicolumn{1}{c}{$\gamma$ ({\rm meV}) } &
   \multicolumn{1}{c}{$T_{K}$ ({\rm mK}) }\\
    \colrule
MTB1 (Fig.~3 in the main text)   & $47.5$ & $-33.0$ & $7.4$ & $14.8$\\
MTB2 (Fig.~2 in the main text)   & $40.5$ & $-18.5$ & $7.7$ & $19.9$\\
\hline 
\end{tabular}
\caption{Anderson model parameters $U,\epsilon$ and $\gamma$ (bare $\gamma_0=\gamma/2$) for the two MTBs investigated, together with the calculated 
  Kondo scales $T_{K}$ from Eq.~(\ref{eq:t0-asymm}). 
}\label{Table1}
\end{ruledtabular}
\end{table}

\section{Numerical renormalization group}
The NRG procedure for the Anderson impurity model consists of the following steps: (i) a logarithmic discretization of the conduction band energies about the
Fermi level ($E_F=0$) $\varepsilon_k \to \epsilon_j, j=\pm 1,\pm 2,\dots$ with $\epsilon_{\pm 1}=\pm D, \epsilon_{\pm j }=\pm D\Lambda^{2-|j|-z}, j=2,\dots$, where $\Lambda>1$ is the discretization parameter and $0<z\le 1$ is the z-averaging parameter \cite{Oliveira1994,Campo2005}, (ii), transformation of the conduction band to a tridiagonal form with new basis states $f_{n\sigma, n=0,1,\dots}$ and hoppings $t_n ~D\Lambda^{-n/2}$. This  leads to the following discrete linear-chain form of the Anderson model,
\begin{align}
 H=\sum_{\sigma}{\epsilon_{\sigma}}n_{\sigma}+Un_{\uparrow}n_{\downarrow}+V\sum_{\sigma}(f^{\dagger}_{0\sigma}d_{\sigma}+ \text{H.c.})+\sum_{n=0}^{\infty}\sum_{\sigma}t_n(f_{n\sigma}^{\dagger}f_{n+1\sigma}+\text{H.c.}).\label{eq:AM-linear-chain-form}
\end{align}
Finally, (iii), Eq.~(\ref{eq:AM-linear-chain-form}) is iteratively diagonalized by using truncated Hamiltonians $H_{N}$ with $N=0,1,\dots$, conduction electron orbitals $f_{n\sigma},n=0,\dots,N$, and the recursion relation $H_{N+1}=H_N+ t_{N}\sum_{\sigma}(f_{N\sigma}^{\dagger}f_{N+1\sigma}+\text{ H.c.}) \equiv {\mathcal T}[H_N]$. This procedure yields the many-body eigenvalues $E_p^N$, eigenvectors $|p\rangle_N$, and also the matrix elements of physical observables of interest, e.g., the matrix elements $\langle p| d_{\sigma}|q\rangle_N$ required for calculating the spin-resolved impurity spectral function. The logarithmic discretization parameter  $\Lambda>1$ separates out the many (infinite) energy scales of the conduction band, from high energies (small $n$) to intermediate energies (intermediate $n$) and low energies ($n \gg 1$), allowing the 
physics to be obtained iteratively on each successive energy scale. The approach is nonperturbative and allows essentially exact calculations for 
local (impurity) thermodynamical and dynamical quantities to be carried out on all temperature and energy scales \cite{Bulla2008}. 
For the calculations in this Letter, we use the full density matrix approach to spectral functions \cite{Hofstetter2000,Anders2005,Peters2006,Weichselbaum2007}.
The spin-resolved spectral functions $A_{\sigma}(E) = -{\rm Im} \left[G_{\sigma}(E)\right]/\pi$ are obtained from the local impurity Green function
$G_{\sigma}(E) \equiv \langle\langle d_{\sigma};d_{\sigma}^{\dagger}\rangle\rangle= (E  - \epsilon_{\sigma} +i\Delta_0 -\Sigma_{\sigma}(E))^{-1}$ via the correlation self-energy $\Sigma_{\sigma}(E)$ \cite{Bulla1998}, where we implement a recent improved representation for $\Sigma_{\sigma}(E)$ \cite{Kugler2022} that guarantees the negativity of the imaginary part of $\Sigma_{\sigma}(E)$, i.e., causality. The spin-resolved differential conductance $(dI/dV)_{\sigma}$ is related to
$A_{\sigma}(E)$ via
\begin{align}
  \left[\frac{dI}{dV}\right]_{\sigma}(eV) & = \int_{-D}^{+D}dE\left(-\frac{\partial f(E-eV)}{\partial E}\right) A_{\sigma}(E),\\
                                 &=  \int_{-D}^{+D}dE\frac{A_{\sigma}(E)}{4k_{B}T \cosh^2((E-eV)/2k_BT)}.\label{eq:NRG-dIdV}
\end{align}
The measured $\mathrm{d}I/\mathrm{d}V$ is related to the spin-resolved spectra by Eq.~(\ref{eq:spin-resolved-dIdV}).
In addition,  we use the z-averaging approach \cite{Oliveira1994,Campo2005} to obtain improved results for spectral functions, differential conductance,  and thermodynamic quantities by averaging over $N_z=192$ realizations of the conduction band using $\Lambda=5$ as in Ref.~\onlinecite{vanEfferen2024}.
For further details of the NRG approach in general we refer the reader to reviews \cite{KGWilson1975, KWW1980a,Bulla2008}.  

\section{Field and temperature dependence of spectral functions and $\mathrm{d}I/\mathrm{d}V$}
\begin{figure}[h]
  \centering 
  \includegraphics[width=0.48\linewidth]{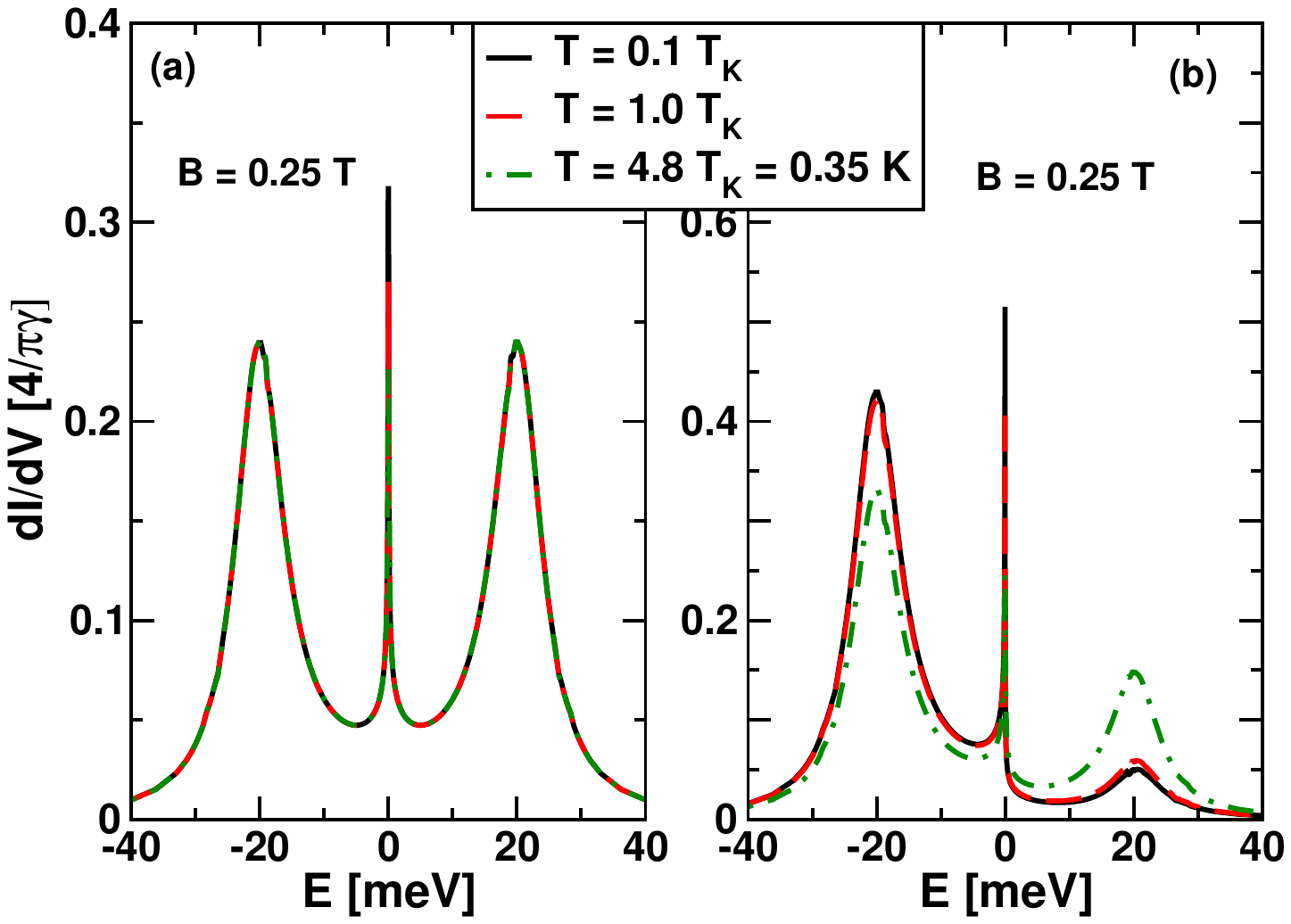}
  \includegraphics[width=0.48\linewidth]{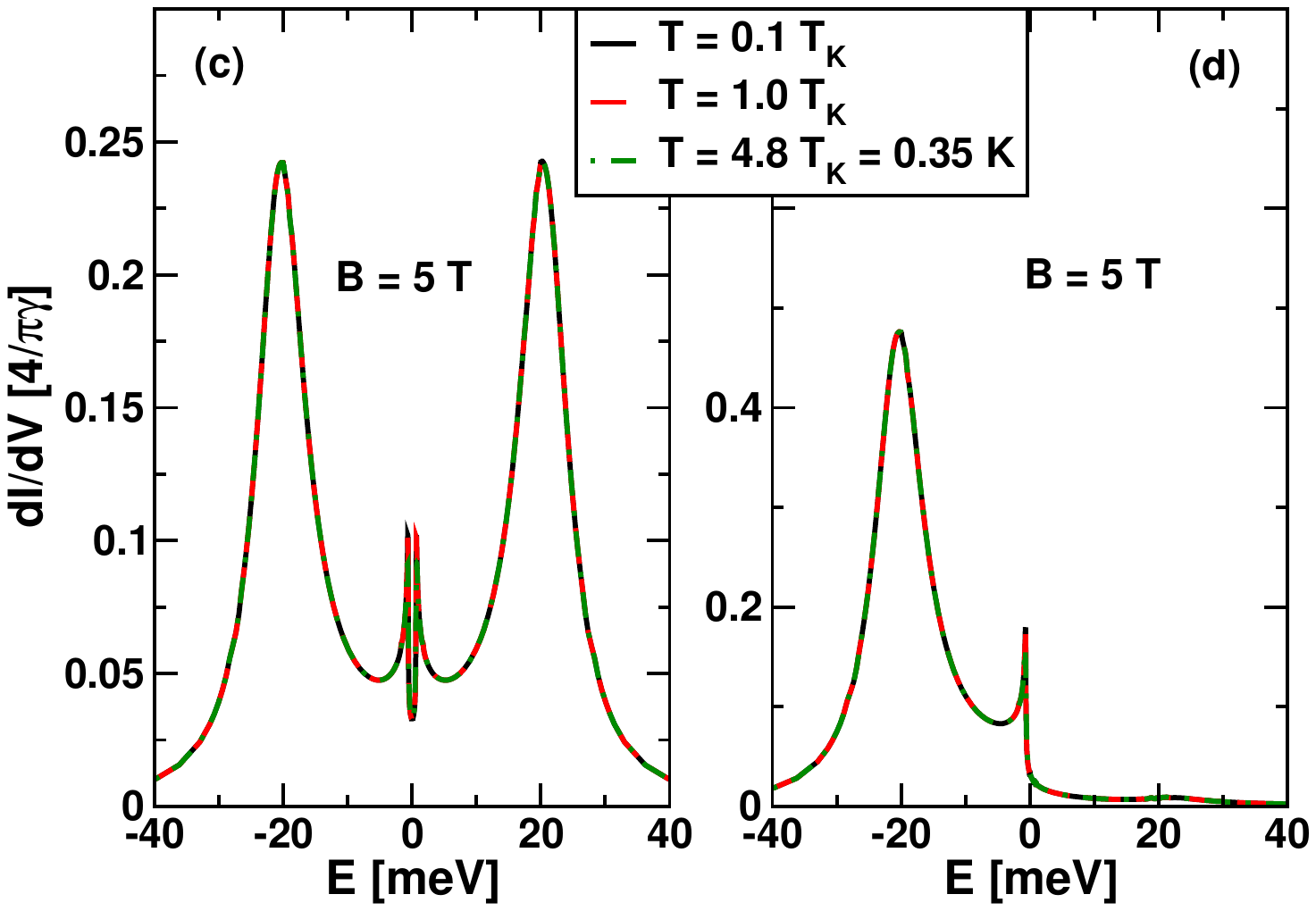}
  \caption{
    (a),(c) Spin-averaged, $\frac{dI}{dV}=\frac{1}{2}(\frac{dI}{dV}_{\uparrow}+\frac{dI}{dV}_{\downarrow})$, and (b),(d) spin-resolved, $\frac{dI}{dV} = \frac{dI}{dV}_{\uparrow}$, spectra for the symmetric Anderson model with $U=40meV, \epsilon=-20meV,\gamma=9meV$. Left panels are for a typical low experimental
    field ($B=0.25 T$ or $g\mu_{B}B=5.8k_BT_K$). Right panels are for a typical high experimental field ($B = 5 T$ or $g\mu_BB=116 k_BT_K$). For each case, the effect of decreasing temperature from above to below the Kondo temperature is shown.
}
\label{fig:spectra-ft-dep}
\end{figure}
 Figure~1 of the main text illustrates schematically the dramatic effect of a magnetic field on the spin-resolved spectral function 
 of the Anderson impurity model. This is shown in more detail in  Fig.~\ref{fig:spectra-ft-dep}, using parameters
 typical for symmetric MTB's: $U=40 meV, \epsilon=-20 meV, \gamma=9 meV$. The spin-averaged and the spin-resolved  $\mathrm{d}I/\mathrm{d}V$ for spin up are shown for  low and high experimental magnetic fields and for temperatures at, $T=0.35$ K, and below, $T=T_K$ and $T=0.1T_K$, the typical experimental temperature. The only significant field and temperature dependence  in the spin-averaged spectral function is in the region of the Kondo resonance [see Figs.~\ref{fig:spectra-ft-dep}(a) and \ref{fig:spectra-ft-dep} (c)]. In stark contrast to this, even a small magnetic field of order $g\mu_B B = k_BT_K$ suffices to cause a large spectral weight rearrangement from the upper excitation at $E=\epsilon+U$ to the lower excitation at $E=\epsilon$ in the spin resolved spectral function [see Figs.~\ref{fig:spectra-ft-dep}(b) and \ref{fig:spectra-ft-dep} (d)]. The effect is largest at low temperatures $k_B T \ll  g \mu_B B$  [Fig.~\ref{fig:spectra-ft-dep}(d)], while for $k_BT$ comparable to or larger than $g\mu_B B$, depolarization of the
 Kondo state starts to take place as indicated in Fig.~\ref{fig:spectra-ft-dep}(b). This remarkable spin-polarization of the Kondo state, with large field induced spectral weight rearrangement, is also reflected in the measured $\mathrm{d}I/\mathrm{d}V$, which probes an unequal ($|p|>0$) admixture of up and down spin spectral functions, see Eq.~(\ref{eq:spin-resolved-dIdV}). 

 \section{Impurity magnetization}
 Detailed information on the polarizability of the Kondo state is contained in the field dependence of the impurity magnetization $m(B)=\frac{1}{2}g\mu_B\langle n_{\uparrow}-n_{\downarrow}\rangle$. This can easily be calculated within the NRG approach as a thermodynamic quantity. We characterize its general behavior in Secs.~\ref{subsec:zeroT+checks}-\ref{subsec:finiteT+checks}, for both zero and finite temperatures, and provide numerical checks on the calculations.
We then discuss in  Sec.~\ref{subsec:magnetization-from-spectra} two approaches, including the one used in the main text, on how to extract the magnetization from the field dependence of the spin-polarized $\mathrm{d}I/\mathrm{d}V$, the quantity directly accessible in STM.

\subsection{$T=0$ magnetization and numerical checks} 
\label{subsec:zeroT+checks}
\begin{figure}[thb!]
  \centering 
    \includegraphics[width=\linewidth]{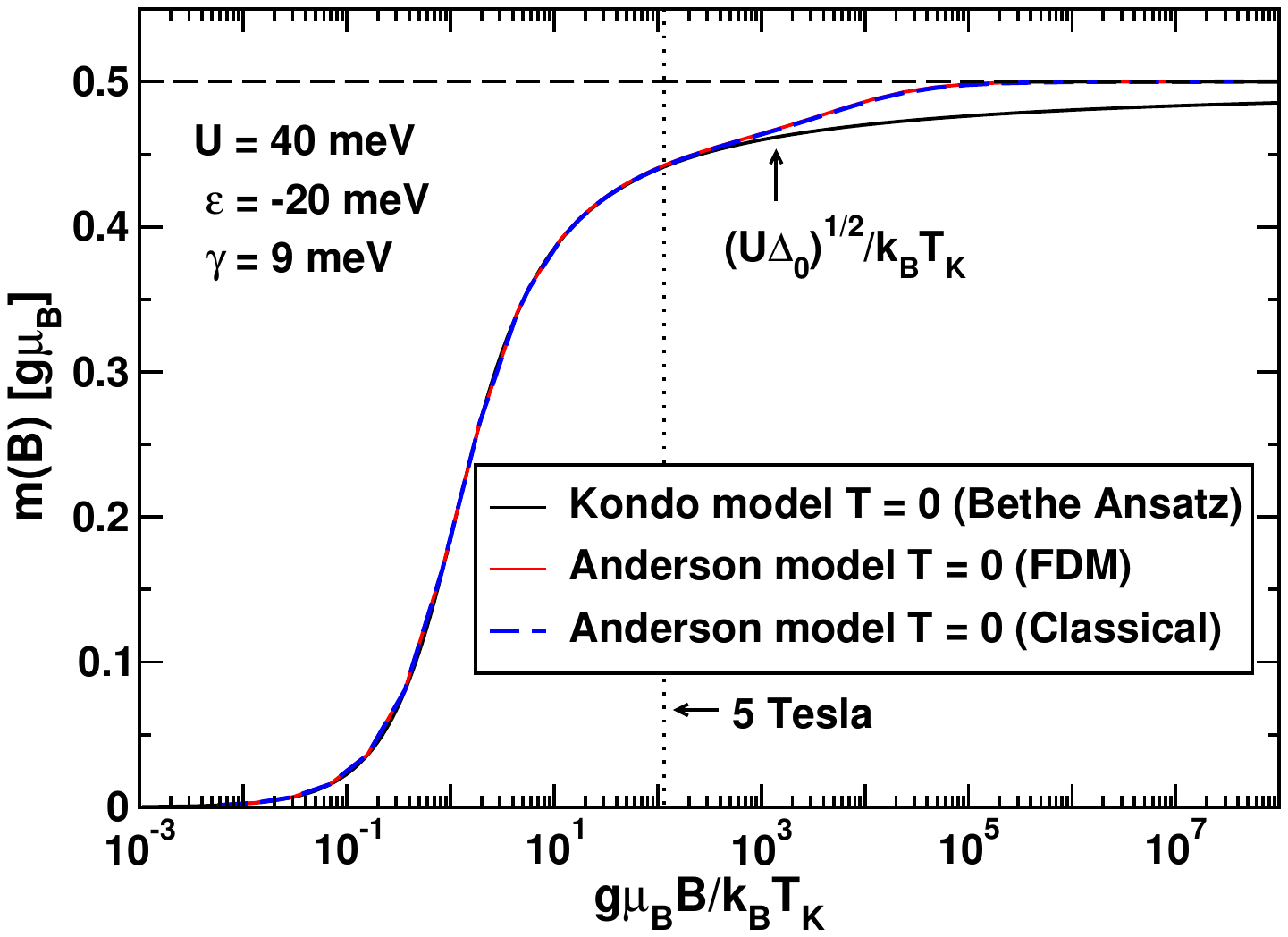}
\caption 
 {Magnetization $m(B)$ versus $g\mu_BB/k_BT_K$ at $T=0$ for the symmetric Anderson model with $U=40$meV, $\epsilon=-20$ meV and $\gamma=9$ meV,  calculated with NRG via full density matrix (FDM) and classical (conventional) approaches. The zero temperature magnetization curve of  the Kondo model (black line, obtained via the Bethe Ansatz) is also shown and is seen to agree with the magnetization of the Anderson model results for fields up to the crossover scale $g\mu_BB_0\sim (U\Delta_0)^{1/2}$ (vertical arrow). Left arrow indicates the maximum experimental field of $5$ Tesla (corresponding to $g\mu_BB/k_BT_K=116$).  
}
\label{fig:MagZeroTemp}
\end{figure}
Figure~\ref{fig:MagZeroTemp} shows $m(B)$ as a function of the dimensionless magnetic field $g\mu_BB/k_BT_K$ for the symmetric Anderson model at $T=0$ for parameters typical for a symmetric MTB. In order to check the NRG calculations, we obtained $m(B)$ within two independent approaches: (i), within the full density matrix (FDM) approach to thermodynamics \cite{Merker2012b}, and, (ii), within the classical (or conventional) approach \cite{KGWilson1975,KWW1980a,Campo2005}. The former uses the complete basis set of discarded states in the NRG \cite{Anders2005,Weichselbaum2007} to evaluate thermodynamic expectation
values $\langle n_{\sigma}\rangle$ and hence $m(B)$, while the latter calculates the impurity contribution to the magnetization as in the original work of Wilson \cite{KGWilson1975}, by first calculating the magnetization of the total system (with field acting both on the electrons and the impurity) and then subtracting from this the contribution from just the conduction electrons alone [see Ref.~\onlinecite{Merker2012b} for further details and comparisons of the two approaches]. Clearly, from Fig~\ref{fig:MagZeroTemp}, one sees that the two approaches give identical magnetization curves for the Anderson model at all fields, thereby providing a rigorous
check on the correctness of the calculated magnetizations.

In Fig.~\ref{fig:MagZeroTemp}, we also show the exact analytical result for the $T=0$ magnetization of the $S=1/2$ Kondo model obtained from the Bethe Ansatz solution  \cite{Andrei1982}:
\begin{equation}
  \frac{m(B)}{g\mu_B}  = \begin{cases}\frac{1}{\pi^{1/2}}\sum_{k=0}^{\infty}\frac{(-1)^{k}}{k!}(k+\frac{1}{2})^{k-1/2}e^{-(k+\frac{1}{2})}\left(c\frac{g\mu_BB}{k_BT_0}\right)^{2k+1}, \;\text{ for $g\mu_BB/k_BT_0<1$},\\
    =  \frac{1}{2}\left[1-\frac{1}{\pi^{3/2}}\int_{0}^{\infty}dt \frac{\sin{\pi t} } {t} e^{-t(\ln t -1)}\Gamma(t+\frac{1}{2})\left(\frac{k_BT_0}{cg\mu_BB}\right)^{2t}\right], \;\text{ for $g\mu_BB/k_BT_0\ge 1$}, 
    \end{cases}
\end{equation}
with $c=\sqrt{e/2\pi}$ and $T_0=\frac{2}{\pi}T_K$. At $T=0$, the magnetization is seen to be a universal function of reduced field ($g\mu_BB/k_BT_K$), i.e., $m(B,T=0)= f_{KM}(g\mu_BB/k_BT_K)$, while at finite temperature it is a universal function of $g\mu_BB/k_BT_K$ and $T/T_K$ \footnote{The universal functions are usually written in terms of $g\mu_BB/k_BT$ and $T/T_K$\cite{Andrei1983} . 
  Setting $T=\alpha T_K$, allows them to be expressed in the manner we stated.}, i.e.,
$$m(B,T)=f_{KM}(g\mu_BB/k_BT_K,T/T_K).$$
From Fig.~\ref{fig:MagZeroTemp},  one sees perfect agreement with the magnetization of the Anderson model for all fields well below the (magnetic field) crossover scale $g\mu_BB_0 \sim (U\Delta_0)^{1/2}$ of the (symmetric) Anderson model \cite{Wiegmann1983a}. Above this scale, the impurity magnetization of the Anderson model deviates from the universal Kondo magnetization curve due to the increasing role of charge fluctuations (see next section). In particular, for $g\mu_BB\gg (U\Delta_0)^{1/2}$,
the Anderson model magnetization rapidly saturates to the fully polarized value, $m(B)/g\mu_B=1/2$, characteristic of a free spin $1/2$. In contrast, such a fully polarized state is impossible to achieve in the Kondo model, since the approach to full polarization $m(B)/g\mu_B=1/2$ is governed by logarithmic corrections \cite{Andrei1982}. This
point is relevant in interpreting the present experiments since the largest fields have $g\mu_BB\ll (U\Delta_0)^{1/2}$ so the magnetization extracted at the largest fields is suppressed by the logarithmic Kondo corrections and should be discernible, at least for the lowest temperature measurements.
\subsection{Finite-temperature magnetization} 
\label{subsec:finiteT+checks}
\begin{figure}[thb!]
  \centering 
  \includegraphics[width=\linewidth]{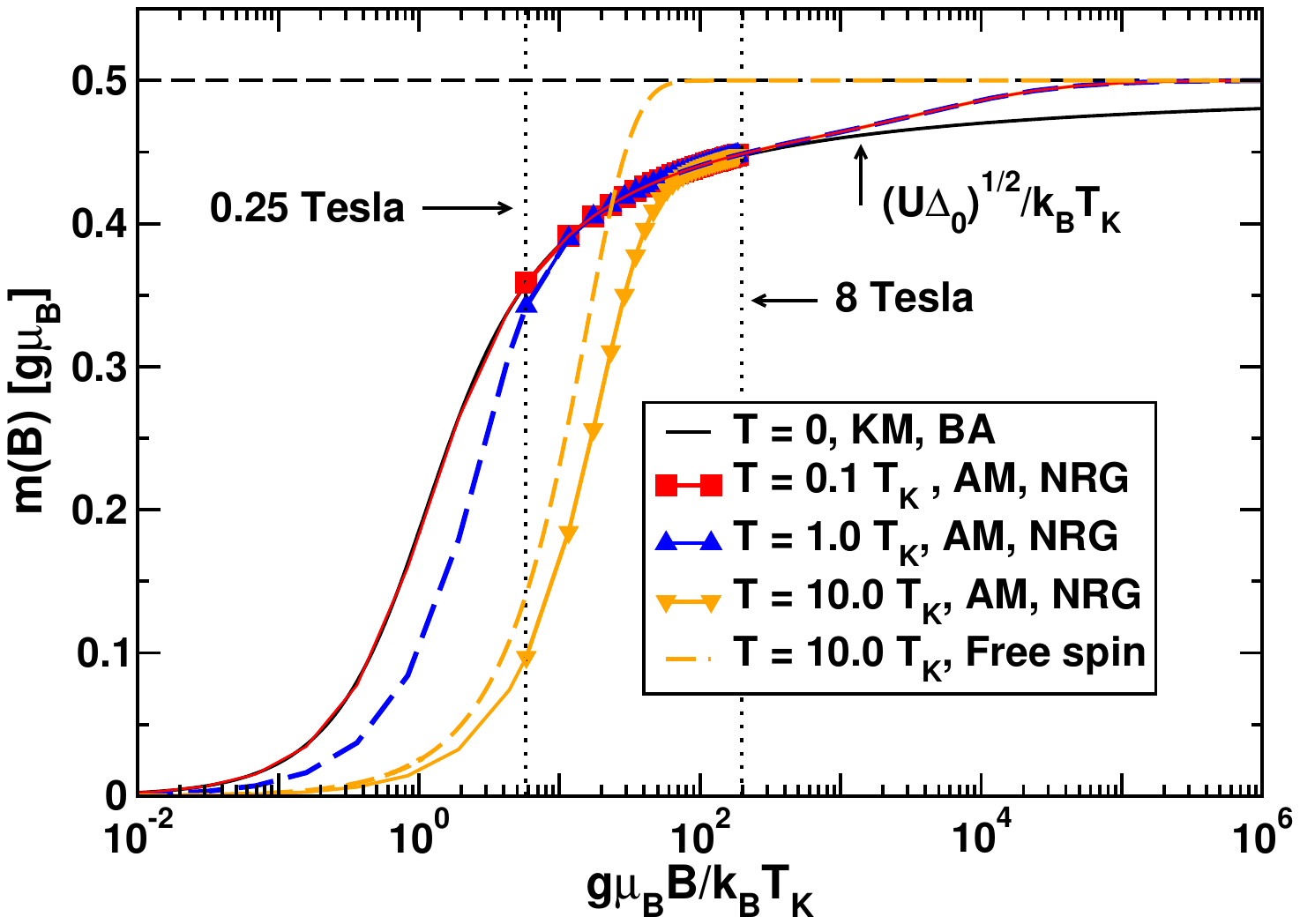}
\caption 
 {Magnetization $m(B)$ versus $g\mu_BB/k_BT_K$ for  the symmetric Anderson model, with the same parameters as in Fig.~\ref{fig:MagZeroTemp} at  finite temperatures: (i) in the strong coupling regime $T=0.1T_K$, (ii), in the intermediate coupling regime $T=1.0 T_K$, and, (iii), in the weak coupling regime $T=10.0T_K$. The $T=0$ magnetization for the Kondo model (via Bethe Ansatz) is also shown for comparison. A typical experimental temperature $T=0.35$ K would correspond to $T\approx 5 T_K$ for the system shown. Symbols  
   indicate the typical experimental field range $0.25-8$ T. While the Anderson model magnetization in the strong coupling regime (i) fits the $T=0$ universal Kondo curve very well on all fields $g\mu_BB \ll (U\Delta_0)^{1/2}$, with increasing temperature, in (ii) and (iii), deviations from this are seen which increase with decreasing field. The free
   spin magnetization, shown for $T=10 T_K$ (orange dashed line), differs markedly from the interacting Anderson model magnetization, except at asymptotically high fields $g\mu_BB \gg (U\Delta_0)^{1/2}$ when both approach the fully polarized value.  
}
\label{fig:MagFiniteTemp}
\end{figure}
For $g\mu_BB$ and $k_BT \ll (U\Delta_0)^{1/2}$, the magnetization of the Anderson model  is a universal function of both $g\mu_BB/k_BT_K$ and $T/T_K$, i.e., $$m(B,T)=f_{AM}(g\mu_BB/k_BT_K, T/T_K),$$ with $f_{AM}(x,y)=f_{KM}(x,y),$
for $x,y\ll (U\Delta_0)^{1/2}$ \cite{Wiegmann1983a} (see  also Sec.~\ref{subsec:universality-mag}).  For $g\mu_BB$ or  $k_BT$ 
comparable to, or larger than the crossover scale $(U\Delta_0)^{1/2}$, the magnetization 
acquires non-universal corrections, i.e.,
it will be dependent on the microscopic parameters (such as $U, \epsilon$ or $\Delta_0=\gamma_0/2$) and $f_{AM}(x,y)$ will deviate from the universal Kondo model magnetization $f_{KM}(x,y)$. 

The effect of temperature on the impurity magnetization of the Anderson model is shown in Fig.~\ref{fig:MagFiniteTemp} for three characteristic 
temperatures in the strong, intermediate and weak coupling regimes, respectively. In the 
%strong coupling regime
former, exemplified by
the case  $T=0.1T_K$, one sees almost prefect coincidence, as is to be expected,  between the  Anderson model magnetization  and the
%universal 
$T=0$ magnetization curve for the Kondo model on all fields $g\mu_B B \ll (U\Delta_0)^{1/2}$. This provides a further check on the NRG calculations
for the Anderson model. Upon increasing temperature through the crossover regime, $T=T_K$, and into the weak coupling regime, $T\gg T_K$, a departure from the $T=0$ Kondo model magnetization is found, also as expected, and starting from low fields. At sufficiently large fields [but still much below $(U\Delta_0)^{1/2}$], the finite temperature magnetization curves for the Anderson model approach asymptotically the $T=0$ Kondo model curve, i.e., the finite temperature Anderson model curve is close to the $T=0$ Kondo model curve for high fields satisfying $(U\Delta_0)^{1/2}\gg g\mu_BB\gg k_BT\gg k_BT_K$,
We can estimate whether signatures of the universal Kondo curve can be observed in the experiment. To this effect, we indicate with symbols in Fig.~\ref{fig:MagFiniteTemp} the experimentally accessible field range. For typical
experimental temperatures, e.g., $T=0.35$ K, corresponding to $T\approx 5 T_K$, intermediate between the two highest temperatures calculated,
Fig.~\ref{fig:MagFiniteTemp} shows that the Kondo curve may well fit a small (high) field range in the measured magnetization curve, with the precise range
depending on the value of $T/T_K$. Thus the slow logarithmic approach of the magnetization to its free spin value at $g\mu_BB\gg k_BT\gg k_BT_K$, characteristic of
the Kondo effect, can be observed in the finite temperature magnetization of the Anderson model. In contrast to this typical Kondo behavior, the 
magnetization for a free spin $1/2$ at $T=10 T_K$ (dashed orange curve) saturates rapidly to the fully polarized value,  thereby
allowing for a clear distinction between the two in experiment. In summary, knowledge of the exact Anderson model magnetization $m(B)$ at any finite temperature,
should allow to fit the measured data over the whole field range and to distinguish this from the free spin behavior.  In the next subsection we discuss how to extract the impurity magnetization from the measured spin-polarized differential conductance.

\subsection{Magnetization from spin-resolved spectra}
\label{subsec:magnetization-from-spectra}

\subsubsection{Magnetization from weight asymmetry of $\mathrm{d}I/\mathrm{d}V$}
\label{subsubsec:weight-asymmetry-approach}
One approach to extracting the impurity magnetization is to consider the field dependence of the ``weight asymmetry'' of the peaks in $\mathrm{d}I/\mathrm{d}V$. By ``weight asymmetry'' we mean the following quantity
   \begin{align}
A_{dI/dV}^w(B) & = \frac{w^{\epsilon+U}(B)-w^{\epsilon}(B)}{w^{\epsilon+U}(B)+w^{\epsilon}(B)}, \label{eq:Asym-weight-definition}
   \end{align}
   where
   \begin{align}
     w^{\epsilon+U}(B) & = \int_{-D}^{+D} (1-f(E))\frac{dI}{dV}(E)dE,\label{eq:w-plus}\\
     w^{\epsilon}(B) & = \int_{-D}^{+D} f(E)\frac{dI}{dV}(E)dE.\label{eq:w-minus}
   \end{align}
The measured (low temperature) $\mathrm{d}I/\mathrm{d}V$, according to Eq.~(\ref{eq:spin-resolved-dIdV-lowTemp}), corresponds to an admixture of up and down spectral functions,
\begin{align}
\frac{dI}{dV}(E=eV) & \approx c^{<,>}\left[\frac{1+p}{2}A_{\uparrow}(E,B)+\frac{1-p}{2}A_{\downarrow}(E,B)\right],\label{eq:polarized-dIdV}
  \end{align}
  with $-1\le p\le +1$ being the polarization of the tip and $c^{<,>}$ are the prefactors defined after 
Eq.~(\ref{eq:spin-resolved-dIdV-lowTemp}).

\begin{figure}[thb!]
  \centering 
  \includegraphics[width=0.8\linewidth]{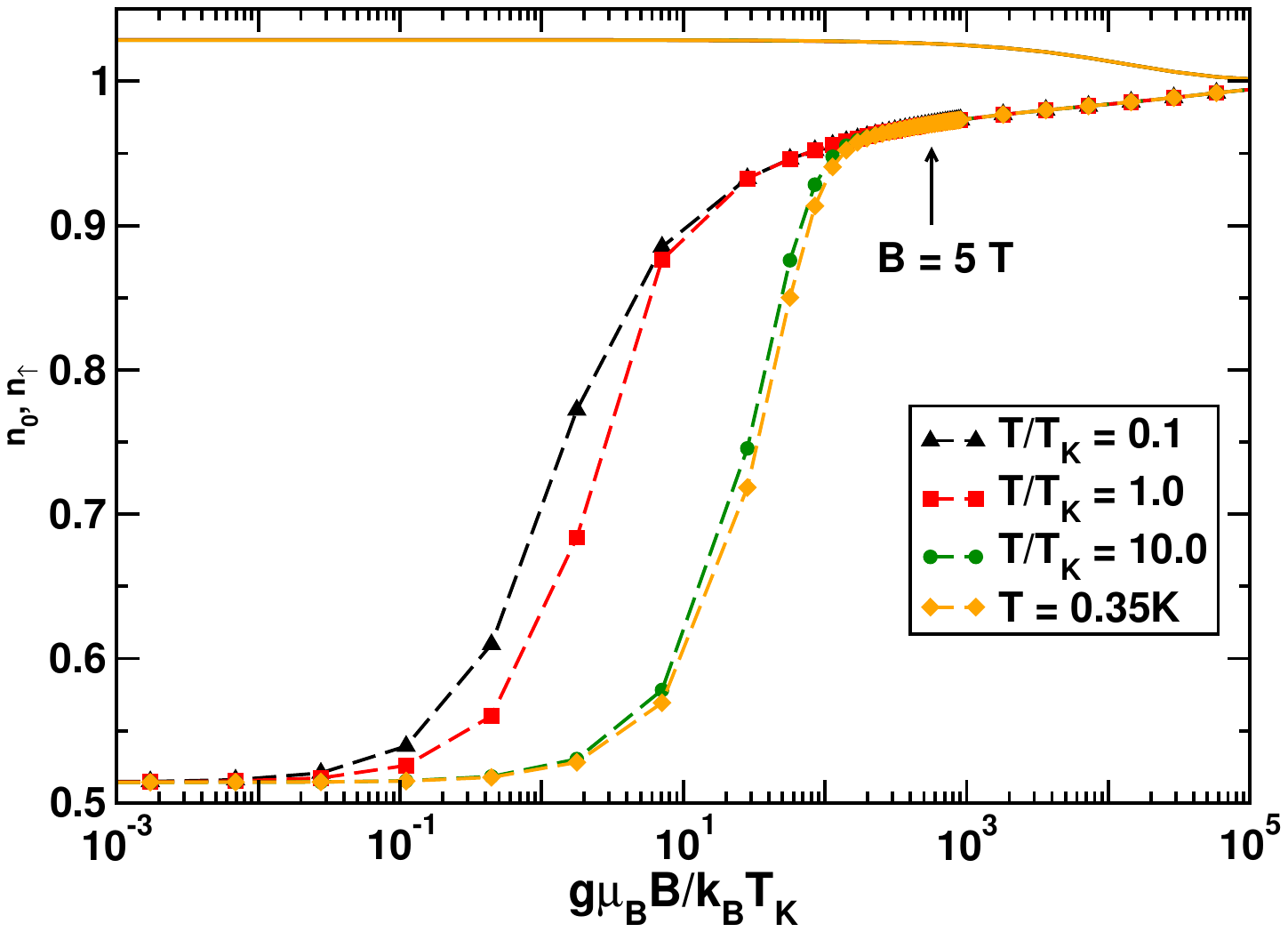}
  \caption{Field dependence of the total occupation $n_0=n_\uparrow+n_\downarrow$ (solid lines) and the up spin occupation $n_\uparrow$ (dashed lines with symbols) for the boundary MTB1 
    with parameters $U=47.5$ meV, $\epsilon=-33$ meV, and, $\gamma=7.4$ meV at several temperatures, including the experimental temperature $T=0.35 K =23.6 T_K$, as well as lower temperatures extending into the strong coupling regime (e.g., $T=0.1T_{K}$).  The experimental fields lie at, and below, 5 T (the vertical arrow indicates a 
field of 8 T). Notice the almost complete absence of field dependence for the total occupation (the four curves at the different temperatures are almost indistinguishable). In contrast $n_{\uparrow}$ varies strongly with field and temperature. 
}
\label{fig:occupations}
\end{figure}

   Substituting (\ref{eq:polarized-dIdV}) into (\ref{eq:w-plus})-(\ref{eq:w-minus}), we find
   \begin{align}
     A_{dI/dV}^w(B) & = 1 - n_0 -2p m(B)/g\mu_B\label{eq:weight-asymmetry-dIdV}\\
     m(B)/g\mu_B & =                      -\frac{A_{dI/dV}^w(B) - A_{dI/dV}^w(0)}{2p}.\label{eq:weight-asymmetry-m}
   \end{align}
   where $n_0=n_0(B)=n_{\uparrow}(B)+n_{\downarrow}(B)$ is the total occupancy of the impurity. While the unknown constants $c^{<,>}$
   cancel in the definition of $A_{dI/dV}^{w}$, the derivation of  (\ref{eq:weight-asymmetry-m}) implicitly assumes that $n_0(B)$ has negligible field dependence. This assumption is obviously true for the symmetric MTB's in the Kondo regime, since $n_0=1$ exactly for all fields by particle-hole symmetry.   However, it holds also for the asymmetric model $\epsilon \neq -U/2$ in the Kondo regime. We demonstrate this for the asymmetric MTB with $U=47.5$ meV, $\epsilon=-33$ meV, and, $\gamma=7.4$ meV   in Fig.~\ref{fig:occupations}. While $n_{\uparrow}$ varies by O(1) in the range $B=0-8$ T, $n_0$ changes by less than $0.1\%$ in the same range and for all temperatures shown [the curves for $n_0$ ($\approx 1$) in Fig.~\ref{fig:occupations} are indistinguishable]. Thus, $n_{0}$ has essentially no magnetic field dependence in the Kondo regime $n_0\approx 1$, a statement that remains true not only at high temperature but also well into the low temperature strong coupling regime ($T\ll T_K$), as exemplified, for example, by the curve for $T/T_K=0.1$ in Fig.~\ref{fig:occupations}. 
This way of extracting $m(B)$ is exact, as applied to the Anderson model. For the experimental system, the evaluation of the weight asymmetry $A_{dI/dV}^w(B)$ from (\ref{eq:Asym-weight-definition}) incurs some errors due to the need to restrict the integration range when evaluating
(\ref{eq:w-plus})-(\ref{eq:w-minus}) in order to avoid contributions from neighbouring levels which arise from Luttinger liquid aspects of the physics of the one dimensional MTB's \cite{Jolie2019}. We shall compare this approach for our MTB's explicitly in the following
section, where we look at a related approach to extracting the magnetization which is based 
%not 
on the field dependence of the peak height asymmetry.
   
   \subsubsection{Magnetization from height asymmetry of $\mathrm{d}I/\mathrm{d}V$}
\label{subsubsec:asymmetry-approach} 
\begin{figure}[h]
  \centering 
  \includegraphics[width=0.8\linewidth]{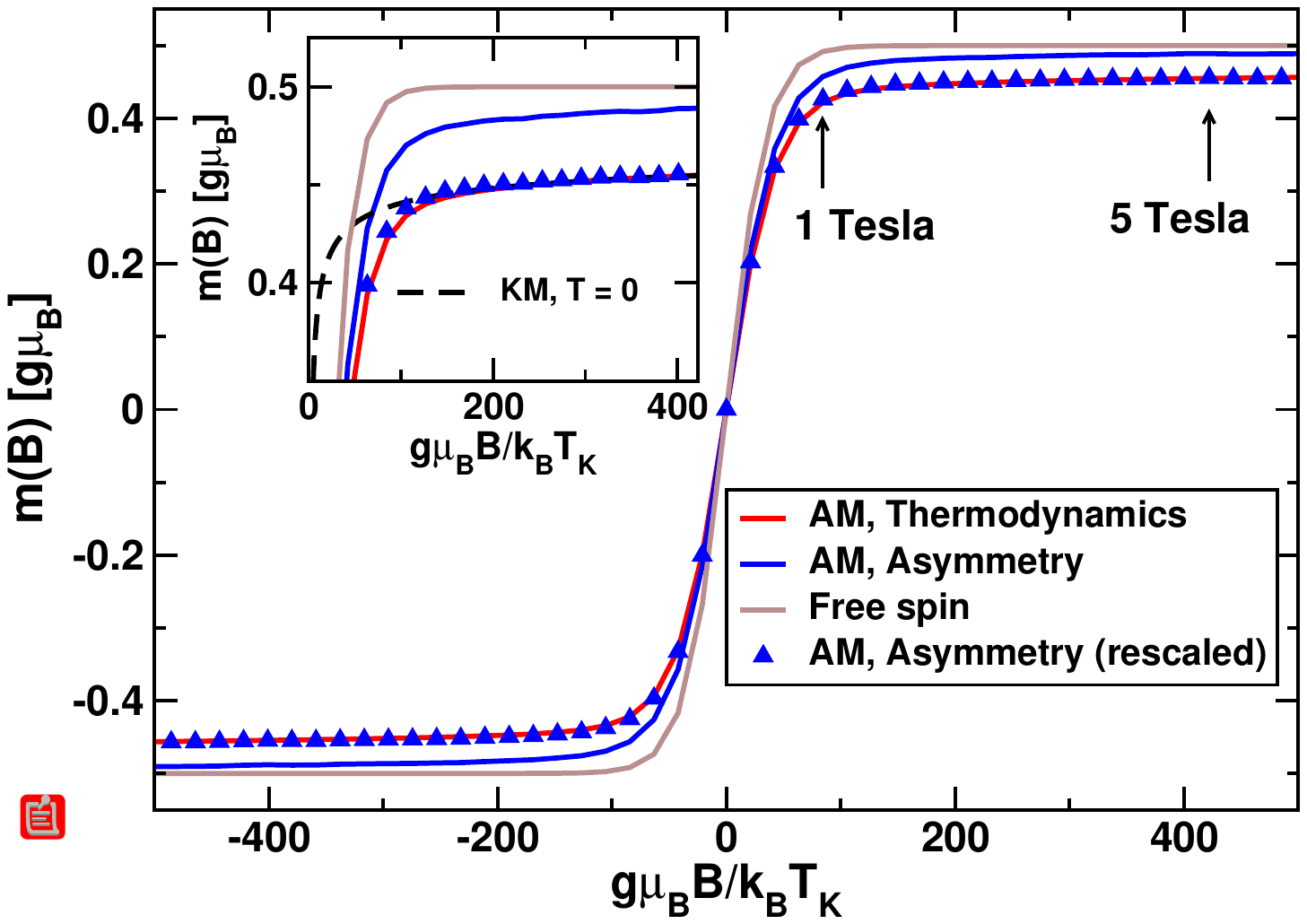}
  \caption{Magnetization $m(B)/g\mu_B$ for the (almost) symmetric boundary MTB2 with Anderson model parameters $U=40.5$ meV, $\epsilon=-18.5$ meV, $\gamma=7.7$ meV at the experimental temperature $T=0.35$ K and field values in the range $-5$ to $+5$ T.  The magnetization from the height asymmetry $m(B)\approx -(A^h_{dI/dV}(B)-A^h_{dI/dV}(0))/2p$ [Eq.~(\ref{eq:height-asymmetry-m})] used $p=0.3$ (blue line) and overshoots the thermodynamic magnetization (red line). Rescaling the height asymmetry via $p\to \tilde{p}\approx 0.322$ results in the blue triangles, which match well the NRG thermodynamic magnetization (red line). The inset shows a more detailed view for $B>0$ and includes the $T=0$ Kondo model magnetization (black dashed line).
}
\label{fig:mag-compare-approaches}
\end{figure}

Another approach to extracting $m(B)$ also uses Eq.~(\ref{eq:polarized-dIdV}) and proceeds via the field dependence of the $dI/dV$ peak heights at $\epsilon$ and $\epsilon+U$, i.e.,
   $h_{\epsilon}(B)=dI/dV_{E=\epsilon}$ and $h_{\epsilon+U}(B)=dI/dV_{E=\epsilon+U}$. The peaks at $E=\epsilon$ and $E=\epsilon+U$ remain Lorentzian in lineshape in the strongly correlated Kondo regime. By contrast the Kondo resonance has a non-Lorentzian lineshape. For a Lorentzian, peak height is proportional to its area. This, together with the fact that in the present experiments, the weight of the Kondo resonance at $E_F$ makes a negligible contribution to the occupation, implies that for spin resolved quantities we have, to a good approximation,$h_{\epsilon,\sigma}\propto n_{\sigma}$ and $h_{\epsilon+U,\sigma}\propto 1-n_{\sigma}$. Thus,  defining the $\mathrm{d}I/\mathrm{d}V$ peak height asymmetry $A_{dI/dV}(B)$ as,
   \begin{align}
A_{dI/dV}^h(B) & = \frac{h_{\epsilon+U}(B)-h_{\epsilon}(B)}{h_{\epsilon+U}(B)+h_{\epsilon}(B)},
     \end{align}
     and noting from Eq.(\ref{eq:polarized-dIdV}) that
\begin{align}
  h_{\epsilon}(B) & \propto  \frac{1+p}{2}n_{\uparrow}+\frac{1-p}{2}n_{\downarrow},\label{eq:h-epsilon}\\
  h_{\epsilon+U}(B) & \propto  \frac{1+p}{2}(1-n_{\uparrow})+\frac{1-p}{2}(1-n_{\downarrow}),\label{eq:h-epsilon+U}
  \end{align}
  we find that,
    \begin{align}
A_{dI/dV}^h(B) & \approx 1-n_0 - 2p\, m(B)/g\mu_B. \label{eq:height-asymmetry-dIdV}
     \end{align}
     As for the weight asymmetry, the unknown prefactors $c^{<,>}$ cancel out in the definition of the height asymmetry. The term $1-n_0$ was shown in Sec.~\ref{subsubsec:weight-asymmetry-approach} to have a negligible field dependence in the Kondo regime. It can be taken as a constant at all fields. Hence, the magnetization can be extracted via
     \begin{align}
       m(B)/g\mu_B & \approx -\frac{A_{dI/dV}^h(B) - A_{dI/dV}^h(0)}{2p}.\label{eq:height-asymmetry-m}
     \end{align}
     While (\ref{eq:height-asymmetry-dIdV}) agrees with the previous expression (\ref{eq:weight-asymmetry-dIdV}) for the ``weight asymmetry'', the height asymmetry
     involves an additional approximation: it assumes that the Kondo resonance near $E_F$ in $A_\sigma$ makes a negligible contribution  $n_{K\sigma}\ll n_{\sigma}$ to the occupation. We shall show by explicit calculation that the error incurred in this approximation can, in part, be taken into account by  renormalizing the parameter $p=\tilde{p}>p$. Figure~\ref{fig:mag-compare-approaches} shows the magnetization calculated from the asymmetry for a tip polarization $p=0.3$, compared with the exact calculation via the thermodynamics for the
     approximately symmetric boundary MTB2 in Table~\ref{Table1}. One sees that the magnetization from the height asymmetry overshoots the thermodynamic value since the actual $p$ used in the above analysis neglects a
     contribution to the polarization of the impurity coming from the Kondo resonance. Upon shifting the asymmetry to ensure $A_{dI/dV}^h(B=0)=0$ and rescaling $p=0.3\to \tilde{p}\approx 0.322$, we see that the asymmetry can be made to coincide with the thermodynamic magnetization to within  a few percent on all field scales relavant to the experiment. We note that this renormalization of $p$ to obtain coincidence between height asymmetry and thermodynamic magnetizations has no physical significance. The tip polarization entering $\mathrm{d}I/\mathrm{d}V$ remains $p$. Instead, the renormalization is required to correct the error made in the height asymmetry approach to the magnetization.
     It reflects the neglect of the states around the Kondo resonance in the contribution to the asymmetry. This error is absent in the weight asymmetry approach  of Sec.~\ref{subsubsec:weight-asymmetry-approach}.
     The inset to Fig.~\ref{fig:mag-compare-approaches} shows a more detailed comparison between the different approaches and includes the $T=0$ magnetization curve for the Kondo model. The different approaches converge to the Kondo curve with increasing field $g\mu_BB\gg k_BT\gg k_BT_K$. 
     The absence of integrations in the height asymmetry approach is a clear advantage over the integration approach. On the other hand, the incurred error described above, requires a rescaling of ``p'' entering Eq.~(\ref{eq:height-asymmetry-m}) in order to match the NRG magnetization.

\begin{figure}[h]
  \centering 
   \includegraphics[width=0.45\linewidth]{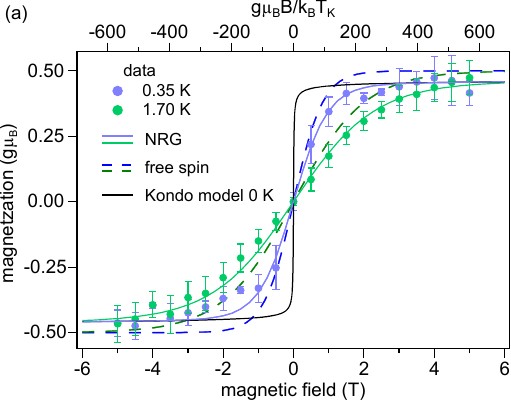}
   \includegraphics[width=0.45\linewidth]{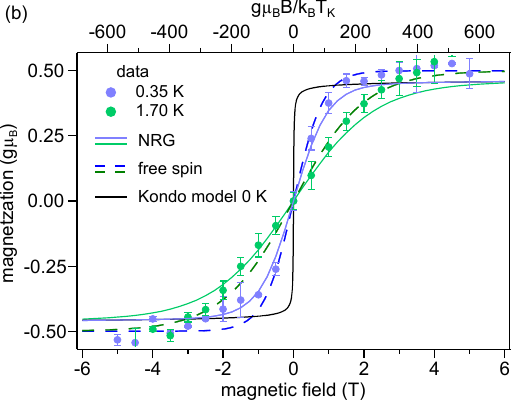}
   \caption{Magnetization $m(B)$ in units of $g\mu_B$ extracted from the field dependence of, (a), the weight, and, (b), the height asymmetry  approaches (using Eqs.~(\ref{eq:weight-asymmetry-m}) and (\ref{eq:height-asymmetry-m}), respectively) for MTB1  
     at temperature $T=0.35$ K ($T_{eff}=0.7$ K) and $T=T_{eff}=1.7$ K. 
     These are compared to the corresponding NRG calculations for the Anderson model at the corresponding temperatures (blue/green solid lines) for MTB1. Also shown is the zero temperature Kondo model magnetization (black solid line). Dashed lines show  the free spin $1/2$ magnetizations. 
     MTB1 Anderson  model parameters: $U=47.5$ meV, $\epsilon=-33$ meV, and, $\gamma=7.4$ meV ($T_K=14.8$ mK).  
}
\label{fig:comparison-weight+height}
\end{figure}

Figure~\ref{fig:comparison-weight+height} shows a comparison of the weight and height asymmetry approaches to the magnetization. Also shown are the NRG thermodynamic magnetization and the free spin $1/2$ magnetization. The weight asymmetry approach gives magnetizations which describe well the Anderson model NRG predictions while they can be clearly distinguished from the free $S=1/2$ magnetization. At asymptotically high fields ($g\mu_B B\gg k_BT\gg k_BT_K$), the thermodynamic Anderson model magnetization and the corrected height asymmetry magnetization approach the %universal 
($T=0$) Kondo model magnetization (black dashed line in the inset). The height asymmetry approach, in the absence of the correction described above, gives magnetizations which are too large, especially at high fields. While this can be taken into account by a rescaling to fit the high field NRG magnetizations, as described above, it is clearly preferable to compare the experimental data with the weight asymmetry approach which
requires no rescaling of the data and which is known to be exact within the Anderson model. This justifies the use of this approach in Fig.~4 of the main text.
   \subsection{Universality of the magnetization }
\label{subsec:universality-mag}
\begin{figure}[h]
  \centering 
   \includegraphics[width=0.8\linewidth]{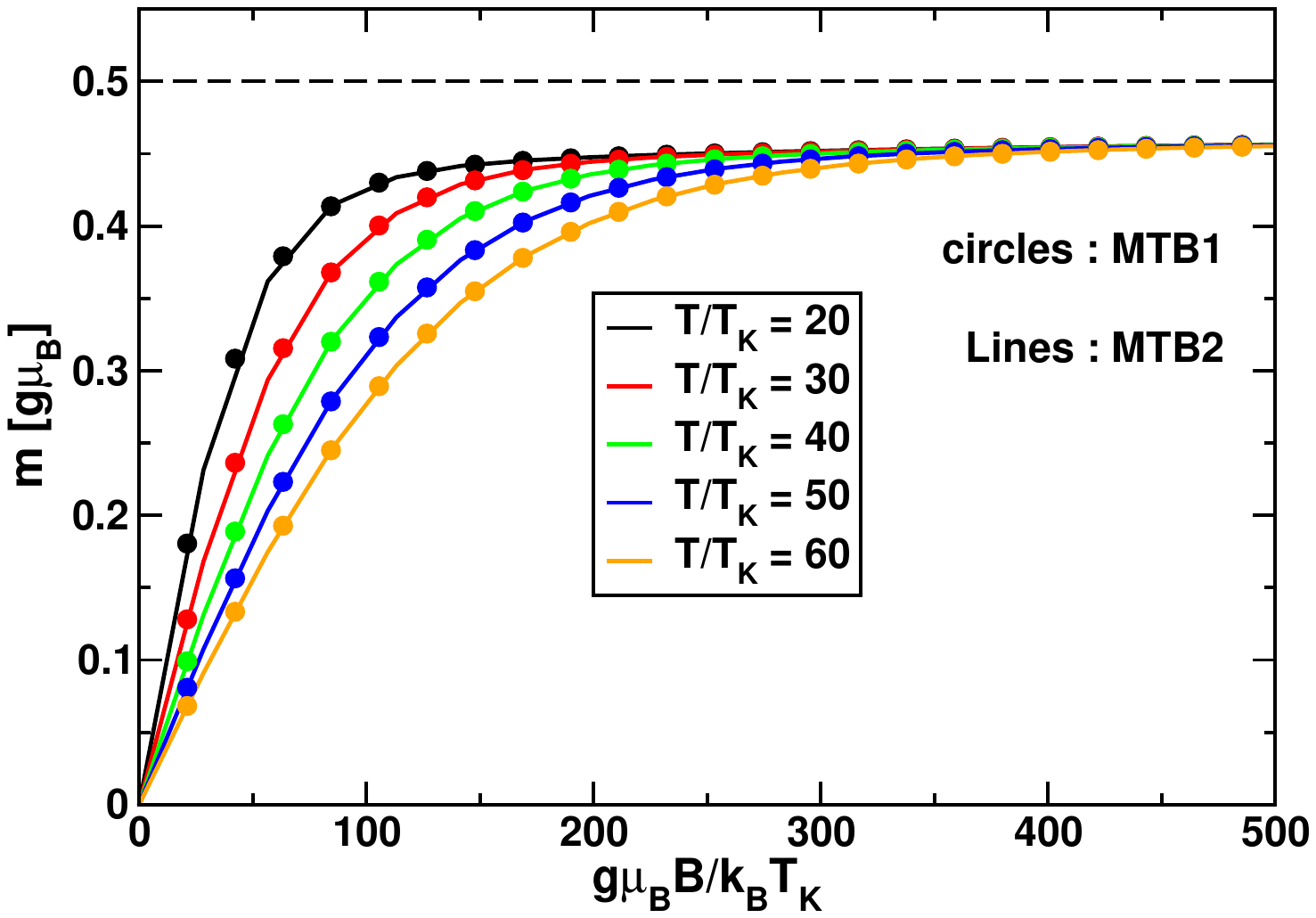}
   \caption{NRG thermodynamic magnetizations $m(B)$ vs $g\mu_BB/k_BT_K$ at fixed $T/T_K$ ratios (given in the legend) for MTB1 (circles) and MTB2 (solid lines). This shows that both MTBs would follow the same universal magnetization curve if their $T/T_K$ ratios were the same.% (and provided that $g\mu_BB\ll (U\Delta_0)^{1/2}$). 
}
\label{fig:mtb1-mtb2-universality}
\end{figure}

As mentioned in Sec.~\ref{subsec:finiteT+checks}, the magnetization of the Anderson model  is a universal function of $g\mu_BB/k_BT_K$ and $T/T_K$,  provided $g\mu_BB$ and $k_BT \ll (U\Delta_0)^{1/2}$. 
In the main text, and in Fig.~\ref{fig:comparison-weight+height}, we have shown the experimentally extracted magnetization for an asymmetric MTB (MTB1)  as a function of $B$ (or $g\mu_BB/k_BT_K$) for each fixed $T$ (equivalently each fixed $T/T_K$) which are compared to the corresponding universal curves from NRG at the same $T$ (or $T/T_K$). The data for MTB2, shown in Fig.~\ref{fig:MTB2-magnetization}, can also be compared to universal NRG curves at the corresponding temperatures $T$. However, since the ratio $T/T_K$ for MTB2 differs from that for MTB1 (due to the differing Kondo scales, see Table~\ref{Table1}), the data for MTB2 will lie, in general, on a different universal curve. To depict both sets of data on the same plot, one would need the same
ratios $T/T_K$, which is not the case for the experimental data.
This explains why we show the data for MTB2 separately in Fig.~\ref{fig:MTB2-magnetization}. Theoretically, one can of course calculate the magnetization vs $g\mu_BBB/k_BT_K$ at fixed $T/T_K$ ratios  for both MTBs. This is shown in Fig.~\ref{fig:mtb1-mtb2-universality}, and illustrates the universality discussed above.

\bibliographystyle{apsrev4-2}
\bibliography{bib_sp-Kondo.bib}